\documentclass[pdflatex, sn-aps, Numbered]{sn-jnl}
\usepackage{graphicx}%
\usepackage{multirow}%
\usepackage{amsmath,amssymb,amsfonts}%
\usepackage{amsthm}%
\usepackage{mathrsfs}%
\usepackage{mathtools}
\usepackage[title]{appendix}%
\usepackage{xcolor}%
\usepackage{textcomp}%
\usepackage{manyfoot}%
\usepackage{booktabs}%
\usepackage{algorithm}%
\usepackage{algorithmicx}%
\usepackage{algpseudocode}%
\usepackage{listings}%
\usepackage{mathrsfs}
\usepackage[cal=rsfso,calscaled=0.9]{mathalfa}
\usepackage[mathscr]{eucal}
\usepackage{relsize}
\usepackage{amsmath}

\usepackage{color,lineno,psfrag,url,cite,verbatim,subfigure,amsfonts,yfonts,dsfont}
\usepackage{xcolor}
\usepackage{lmodern}
\usepackage[T1]{fontenc}
\usepackage[utf8]{inputenc}
\usepackage{soul}
\usepackage{bm}
\usepackage{arydshln}
\definecolor{a}{rgb}{0.1,0.2,0.6}  
\definecolor{v}{rgb}{0.1,0.5,0.3}
\definecolor{b}{rgb}{0.8,0.3,0.3}
\usepackage{nicefrac,xfrac} 
\DeclareFontFamily{OT1}{pzc}{}
\DeclareFontShape{OT1}{pzc}{m}{it}{<-> s * [1.10] pzcmi7t}{}
\DeclareMathAlphabet{\mathpzc}{OT1}{pzc}{m}{it}
\newcommand{\OP}[1]{\scalebox{1.2}{$\mathcal{#1}$}}  
\newcommand{\OQ}[1]{\scalebox{1.2}{$\mathscr{#1}$}}  

\usepackage{upgreek}

\begin{document}

\title[Westervelt derivation]{From deformation theory to a generalized Westervelt equation}


\author[1,2]{\fnm{Mariano} \sur{Caruso}}\email{mcaruso@ugr.es}
\equalcont{These authors contributed equally to this work.}

\author[4,5,6,7]{\fnm{Guillermo} \sur{Rus}}\email{grus@ugr.es}
\equalcont{These authors contributed equally to this work.}

\author*[3,5,6,7]{\fnm{Juan} \sur{Melchor}}\email{jmelchor@ugr.es}
\equalcont{These authors contributed equally to this work.}

\affil[1]{\orgdiv{\texttt{FIDESOL}}, \orgname{Research Center}, \orgaddress{Granada}}

\affil[2]{\orgdiv{\texttt{UNIR}}, \orgname{Universidad Internacional de La Rioja}, \orgaddress{Logroño}}

\affil[3]{\orgdiv{Department} of \orgname{Statistics and Operations Research}, \orgaddress{\street{University of Granada, Spain}}}

\affil[4]{\orgdiv{Department} \orgname{of Structural Mechanics, University of Granada, Spain}}

\affil[5]{\orgdiv{Instituto} \orgname{de Investigaci\'on Biosanitaria, ibs.GRANADA}, \orgaddress{\street{18012 Granada, Spain}}}

\affil[6]{\orgdiv{Research Unit}, \orgname{ ``Modelling Nature'' (MNat)}, \orgaddress{\street{Universidad de Granada, 18071 Granada, Spain}}}
\affil[7]{\orgdiv{Laboratory}, \orgname{Ultrasonics, University of Granada, Spain}}


\abstract{The Westervelt equation describes the propagation of pressure waves in continuous nonlinear and, eventually, diffusive media. The classical framework of this equation corresponds to fluid dynamics theory. This work seeks to connect this equation with the theory of deformations, considering the propagation of mechanical waves in nonlinear and loss-energy media. A deep understanding of pressure wave propagation beyond fluid dynamics it is required to be applied to medical diagnosis and therapeutic treatment.
A deduction of a nonlinear partial differential equation for pressure waves is performed from first principles of deformation theory.  The nonlinear propagation of pressure waves in tissue produces high-frequency components that are absorbed differently by the tissue, thus, distinguishing each of these modes is essential. An extension of the Westervelt equation beyond fluids media is required. In order to include the behaviour of any order harmonics, a generalization of this equation is also developed.}




\maketitle

\section{Introduction}\label{sec1}

The Westervelt equation is a nonlinear partial differential equation, that describes the propagation of pressure waves with different interaction with the media: exhibit a nonlinear behaviour and eventually energy loss. Some of these ideas, principally the nonlinearity, were derived first by P. Westervelt \citep{Westervelt63} from the fluid dynamics theory.

From a mathematical point of view, in a linear wave equation a sum of two independent solution is also a solution of this equation. In general a linear combination, or \textit{superposition}, of separate  monochromatic waves, i.e. with a definite wavelength, is another solution of a linear wave equation. Also, there is no interaction between any two waves with different wavelength. In other words, each of these waves is propagated independently without distortion, i.e. a propagation without change of wave shape. On the other hand, in a nonlinear partial differential equation it is not generally possible to combine known solutions into new solutions, a superposition of any two independent solutions of this equation is not generally a solution of this equation \citep{landau86, Marsden}. The nonlinearity is a mathematical property of the equations, however, under the assumption of that a  concrete nonlinear equation describes the behaviour of a given system, in the present case the propagation of waves through the media, the nonlinearity is also a property of the interaction itself. 

These nonlinear properties was included by  P. Westervelt from a quadratic term in the pressure \citep{Westervelt63}. Westervelt's work was subsequently improved including an energy dissipation term due to the interaction between the medium and the pressure waves, and is described by a diffusive term in the differential equation  \citep{hamilton1998nonlinear,Jimenez,SHEVCHENKO2015200}.
Although these improvements were not made by Westervelt himself, the nonlinear differential equation that contains a quadratic pressure terms and a dissipative term defined by a temporal derivative of order 3 is called by extension a \textit{Westervelt equation}.
 
Note that this kind of partial differential equations has been used for medical purposes as a basic mathematical model, e.g. the correct assessment of the ultrasound dose required for the medical treatment. In addition this kind of equations has been used for medical and industrial  applications, such as diagnostic ultrasound  \citep{Pernot02,Simon98,Floch97}, thermotherapy of tumors \citep{Connor02,Hallaj01,Floch99}, lithotripsy \citep{clason2009boundary, taraldsen2001generalized, Averkiou99}, ultrasound cleaning and sonochemistry  \citep{Kaltenbacher07,Dreyer00}, Fast Ultrasound Image Simulation (FUIS) \citep{karamalis2010fast, varray2011simulation} that can simulate realistic ultrasound images in a short time, High-Intensity Focus Ultrasound (HIFU) \citep{solovchuk2012effects, xiaorui2011simulation, thiriet2014hifu}, etc. Linear models are not applicable for high intensity waves presents in the protocols of the aforementioned applications, simply are not valid in these regimes of wave propagation. The appearing nonlinear effects requires a more sophisticated mathematical models, given in principle by nonlinear differential equations. Although the Westervelt equation has been used for these applications, but it should not be forgotten that this was deduced from fluid dynamics. 

Essentially the Westervelt equation or similar equations are applied but outside the context of fluid dynamics. There is no derivation anchored to the foundations of the theory of deformations in continuous media. On the other hand, in the first derivation of Westervelt equation \citep{Westervelt63} a lossless energy context was considered, i.e. the viscous and thermal conduction effects were neglected. Nevertheless it is possible to complete this deduction adding viscous and thermal conduction effects,  but always in the fluid dynamics. For a modern review see  \citep{hamilton1998nonlinear,Jimenez,SHEVCHENKO2015200}. 
In section \ref{westervelt habitat}, a briefly description for Westervelt equation is presented from fluid dynamics, which is its classical \textit{habitat}. 

In the deformation theory context, if the changes in size and shape are not sufficiently small then the linear theory of deformation must be completed through the nonlinear terms. 
The general equations of deformation theory are presented for \textit{quasi-static} and \textit{irreversible} regimes, and a comment about the solutions of mechanical waves propagation using perturbation theory is detailed in section \ref{nonlinear continuous media}.

There is a connection between the perturbative method applied to the nonlinear differential equation with its harmonics. Inasmuch as that the differential equation is composed by the sum of a linear and a nonlinear, under the condition that the first dominate over the last, then the nonlinear terms can be considered as a perturbation of a linear equation. This allows to express the solution as a sum of functions, that depending on the required precision will be finite or countable infinite quantity of that terms. The first term of this sum is a solution of the linear wave equation, namely the \textit{fundamental} solution. Each of the next terms is a modulation in amplitude and phase of the \textit{fundamental} term. The frequency of the $n-$term is the $n-$multiple of the frequency of the fundamental term, for $n>1$ each term of this sum is a higher order harmonic of the fundamental \citep{hamilton1998nonlinear,Bochud}.  Also, there is a relation between the amplitudes of this terms with the parameters that characterize the propagation media  \citep{hamilton1998nonlinear,Bochud,ma12040607}. This paragraph will be more clear in section \ref{nonlinear continuous media}, but at this moment it is only necessary to notice that a measurement of the amplitudes of these harmonics thus provide an information about the propagation medium. In other words, the solution of the nonlinear theory written in terms of harmonics of the fundamental solution (which corresponds to the linear theory) allows to characterize the materials on which the wave propagates.

The nonlinear wave propagation is particularized to one dimensional $P-$wave equation to extract nonlinearity parameter of first order, namely $\beta$, which has been widely used in the literature  \citep{hamilton1998nonlinear,guyer99b}. The next to leading order parameters beyond $\beta$ are also considered. The aim of this paper is not only perform a deduction of the Westervelt equation from the theory of deformations in continuous media, but  also generalize this equation in order to include high order nonlinear terms, in correspondence with high order harmonics, in section \ref{Westervelt from Elasticity}. The generalized Westervelt equation define an ideal environment to develop future applications to  promote technological progress in the field of tissue ultrasound mechanics. For example, regarding the propagation of pressure waves in tissue the highest-frequency components waves are absorbed more rapidly \citep{GIAMMARINARO2018236}.  This different behaviour, due to the frequency response of the media, shows that the study of each harmonic must be included. A deeper understanding of the nonlinear behaviour of the media must be achieved in order to find new technological applications and improve those that already exist.
For example, signal modeling has also been proven to be an useful tool to characterize damage materials under ultrasonic evaluation \citep{Bochud,donskoy2,Sutin1995,Zarembo1989}. This will serve as a \textit{cyclical motivation} since enriching the models will lead to a more complex analysis of these signals and perhaps a greater content of information. From the theoretical point of view, the road is to build robust models that describe and characterize the properties and the damage of materials, on this subject will treat the following sections.

Understanding the physical world and its phenomena is crucial for advancing scientific knowledge and developing practical applications that can improve our lives. By deriving equations from first principles of physics, we can gain a deeper understanding of the underlying mechanisms that govern the behavior of natural systems. This not only validates the accuracy of the equations but also enables us to generalize their use to a wide range of related phenomena. The rigor of the scientific process ensures that our knowledge is grounded in evidence and not just speculation. Finally, the development of new technologies that stem from these equations can have a significant impact on many areas of society, including medicine, engineering, and more. Therefore, by pursuing these points, we can contribute to the advancement of science and the betterment of humanity.

\section{Nonlinear behaviour in fluids and quasifluids dynamics} \label{westervelt habitat}

The nonlinear behaviour systems can be \textit{thermodynamically} understood from a state equation: a mathematical expression that involves a certain number of thermodynamic variables, e.g. pressure, temperature, volume, density, entropy, among others. Without loss of generality the state equation is given implicitly as a surface $F(\varrho,S,P)=0$, in terms  of the density $\varrho$,  
entropy $S$ and pressure $P$. If this state equation can be written explicitly as a function of density, and temperature: 
 $P=\varphi(\varrho,S)$. 
It is possible to considered a Taylor series of the mentioned state equation, around $\varrho_0$ and for a fixed entropy value $S$, given by
\begin{equation}\label{pressure}
P=\sum_{n \geq 0} \frac{1}{n!}\,(\varrho-\varrho_0)^n\,\partial^{(n)}_\varrho \varphi(\varrho_0,S),
\end{equation}
Considering the following \textit{pressure  parameters}
\begin{equation}\label{A,B,C...}
A\coloneqq \varrho_0\,\partial_\varrho  \varphi(\varrho_0,S),\quad B\coloneqq 
\varrho_0^2\,\partial^{2}_\varrho\varphi(\varrho_0,S),\quad C\coloneqq \varrho_0^3\,\partial^{3}_\varrho \varphi(\varrho_0,S),\; \cdots
\end{equation}
where $\varrho_0$ is the  unperturbed  density value,  denoted by $P_0=\varphi(\varrho_0,S)$ and $c_0^2:=\partial_\varrho \varphi(\varrho_0,S)$, $c_0$ is the small signal sound speed.  The expression \eqref{pressure} is rewritten as \citep{hamilton1998nonlinear}
\begin{equation}\label{under pressure}
p=A\left( \frac{\rho}{\varrho_{0}} \right)+\frac{B}{2!}\left( \frac{\rho}{\varrho_{0}} \right)^{2}+\frac{C}{3!}\left(\frac{\rho}{\varrho_{0}} \right)^{3}+\cdots,
\end{equation}
where $p$ and $\rho$ can be interpreted as the perturbation of the pressure and density $P_0$ and $\varrho_0$, respectively, because $P=P_0+p$ and $\varrho=\varrho_0+\rho$. The expression \eqref{under pressure} retaining up to order $2$ in $\rho$ is
\begin{equation}\label{sound pressure}
p\simeq c^2_0 \rho + \frac{B}{2A}\frac{c_0^2}{\varrho_0}\rho^2
\end{equation}
Note that the dimensionless quantity $B/A$ is the first parameter that quantifies nonlinear effects in $p$ and is  usually calculated for selected fluids, liquefied gases   \citep{hamilton1998nonlinear,Jimenez,SHEVCHENKO2015200}. 

A state equation  does not introduce a dynamic, i.e. does not pronounce on the way in which the system can evolve between two given states. Strictly speaking, a state equation is thermodynamically valid if the system is in an equilibrium state. However, it is possible to deduce an equation that includes the time evolution of such states, without formally addressing the study of non-equilibrium thermodynamics, the main assumption is that the system is always in a state given or described by the state equation. Under the hypothesis that the state equation is satisfied in each time, a formal way to introduce the dynamics of such states is to consider the conservation of mass in terms of a continuity equation for the density, the momentum conservation in terms of the Navier-Stokes equation for the velocity field that includes, eventually, the viscous forces, and finally the Fourier equation that describe the heat-transfer effects due to the thermal conduction behaviour of the media.  


In 1963 P. Westervelt deduces an equation that describes the propagation of pressure waves in nonlinear media from Lighthill's  \citep{Westervelt63}. This work did not include the possibility of loss energy. Nevertheless a modern review of the Westervelt equation can be performed from the state equation \eqref{sound pressure}, the continuity equation and the Navier-Stokes equation, and describes the pressure waves propagation as
\begin{equation}\label{WesterveltEq}
\frac{\partial^2 p}{\partial t^2} -c_0^2 \nabla^2 p =\frac{\beta}{\varrho_0 c_{0}^{2}}\;\frac{\partial^2 p^2}{\partial t^2}+\frac{\delta}{c_{0}^{2}}\,\frac{\partial^3 p}{\partial t^3}
\end{equation}
where $p$ is the pressure that depends on the spatial coordinates and time. The left-hand side of the equation \eqref{WesterveltEq} contains the terms of a linear wave equation. The right-hand side contains the nonlinear terms, characterized by the coefficient $\beta$, which is obtained from the expression \eqref{sound pressure} and is given by
\begin{equation}
\beta=1+\frac{B}{2A}.
\end{equation}
The coefficient $\delta$, namely the sound diffusivity, characterize the energy loss, which comes from two different interaction with the media: viscosity and thermal conduction. The sound diffusivity $\delta$ is given by 
\begin{equation}\label{deltadeWestervelt}
\delta =\frac{1}{\varrho_0}\Big(\frac{\,4\,}{3}\eta +\zeta\Big)+\frac{k}{\varrho_0}\Big(\frac{1}{c_{_{V}}}-\frac{1}{\,c_p\,}\Big),
\end{equation}
where $\eta$  is the shear viscosity, $\zeta$ is the bulk viscosity,  $ k$ is the thermal conductivity, $c_{_V}$ and $c_p$ are the specific heat at constant volume and pressure respectively. For a complete and modern deduction of the Westervelt equation in the fluid dynamics context see \citep{hamilton1998nonlinear,Jimenez,SHEVCHENKO2015200}.

One of the ideas of this work is not only recover the Westervelt equation \eqref{WesterveltEq} in the context of deformation theory, but also promote a formulation capable of describing and identifying the effects of higher order harmonics. The nonlinear terms presents in that equation requires the study of propagation of nonlinear waves in deformable bodies as a counterpart. The wave diffusion associated to energy loss of the system involves the incorporation of viscous and thermal conduction effects between the wave propagation and the media. Therefore, the study of deformable bodies which nonlinear behaviour and energy loss will be the propitious scenario to reconstruct the dynamic dictated by the Westervelt equation.

\section{Nonlinear behaviour  in deformable solids} \label{nonlinear continuous media}

In this section a brief overview of the classical theory of deformation of continuous media is given. In order to avoid ambiguities the term \textit{de-formation} includes material changes in \textit{size} and \textit{shape}, not only in \textit{form}.

Regarding the notation of tensor quantities, including vectors, we denote its components in normal typographical style specifying its indices. While the tensor quantities as a whole, will be shown in bold typeface.   

Beginning with the \textit{displacement field}, i.e. a vector field $\pmb{u}=(u_1,u_2,u_3)$ defined from that undeformed $\{x_i\}_i$ and deformed $\{x'_i\}_i$ spatial coordinates as $u_i=x'_i-x_i$, where $i\in \{1,2,3\}$. Then the \textit{strain tensor} is introduced and represented by a second order tensor field $\pmb{\xi}$, which is related with $\pmb{u}$ according to 
\begin{equation}\label{strain}
\xi_{ij}=\frac{1}{2}\big(u_{i,j}+u_{j,i}+u_{k,i}u_{k,j}\big),
\end{equation}
where $u_{i,j}:=\partial_{x_j} u_i$, $ \, i,j \in \{1,2,3\}$ and the spatial location of each point of the medium is represented by cartesian coordinates $(x_1,x_2,x_3)$. Both tensors $\pmb{u}$ and $\pmb{\xi}$ depends on spatial variables $\pmb{x}=(x_1,x_2,x_3)$, namely  \textit{material} or \textit{lagrangian} coordinates, and time $t$. 

On the other hand, the internal forces of the material can be represented from another second order tensor field $\pmb{\sigma}$ that, also, depends on spatial variables $\pmb{x}$ and time $t$, namely total \textit{stress tensor}. The element $\sigma_{ij}$ of the stress tensor  $\pmb{\sigma}$ is the $i-$component of the sum of the involved forces per unit area perpendicular to the $x_j-$axis. 

The dynamics of the system is regulated through the equation
\begin{equation}\label{dynamics}\rho\,\partial_{tt} \pmb{u}(\pmb{x},t)=div\,\pmb{\sigma}(\pmb{x},t),\end{equation}
where $\rho$ is the density of non-deformed body and $(div\,\pmb{\sigma})_i=\partial_{x_j}\sigma_{ij}$, the \textit{summation convention} is used, i.e. a summation on repeated indices is understood. The $i-$component of the force exerted by the region $\Omega$ on its surrounding parts is given by $\int_\Omega (\partial_{x_j}\sigma_{ij})\, dV $.  
Reciprocally, the force which the surrounding parts of $\Omega$ exerts on the region $\Omega$ it is the same with the opposite sign. \label{signo fuerzas} This idea it will be useful to define pressure.

In this work only two kind of forces are considered: \textit{elastic} and \textit{dissipative}, and  there is a stress tensor associated to each one of these: the \textit{elastic stress tensor} $\pmb{\sigma}$ and the \textit{dissipative stress tensor} $\pmb{\sigma}'$, respectively . There is a functional relation between the strain tensor $\pmb{\xi}$  and the elastic stress tensor $\pmb{\sigma}$, called \textit{constitutive law}. On the other hand, there is also a functional relation between the partial derivative of the strain tensor  $\partial_t\pmb{\xi}$ and the dissipative stress tensor  $\pmb{\sigma}'$. The explicit form of these tensorial relation depends on the nature of the process. According to thermodynamics, two types are essentially distinguished: \textit{quasi-static} process and \textit{irreversible} process. 

There are different types of forces associated with these processes. In the first case we have conservative-type forces, which are derived from an elastic potential. While in the second type of processes the possibility of energy exchange with the media is incorporated, this will give rise to dissipative forces, which can not be derived from the same type of potential. We will use the formalism of the dissipation function to derive these forces \citep{landau80}. In summary, the total stress tensor  for the first of these processes is $\pmb{\sigma}$, while for the second type of these processes is $\pmb{\sigma}+\pmb{\sigma}'$. In the followings subsections, the aforementioned processes and the forces involved in each of these will be specified.

\subsection{Quasi$-$static deformation process}\label{quasistatic process}

In this section, the involved velocities in the deformation process are considered infinitesimal or negligible. The system evolves slowly and in each time can be considered in a state of thermodynamic equilibrium. In other words, the deformation is considered as a \textit{quasi-static} process.

On the other hand a reversible process is also a quasi-static one. But the reciprocal is not true, because a quasi-static process involving entropy production is not reversible. In practice, a quasi-static process can approach to a reversible process, as much as possible. The energy of the body is conserved, then all the forces come from a potential energy per volume unit, $U$. The involved \textit{elastic} forces can be written as a tensor,  namely \textit{elastic stress tensor}, denoted by $\pmb{\sigma}$. A thermodynamic analysis,  including the nonlinear case or large deformations in the strain tensor \eqref{strain}, reveals that a functional relation between the strain tensor $\pmb{\xi}$ and the  elastic stress tensor  $\pmb{\sigma}$ that requires a concrete potential energy $U$ \citep{landau86} given by 
\begin{equation}\label{first stress}
\sigma_{ij}=\frac{\partial \,U}{\partial u_{i,j}},
\end{equation} 
In many situations, the potential energy $U$ is given as a function of the strain tensor $\pmb{\xi}$, using \eqref{strain} and the chain rule
\begin{equation}
\frac{\partial \,U}{\partial u_{i,j}}=\mathlarger{\mathlarger{\sum}}_{k,l}\;\frac{\partial \,U}{\partial \xi_{kl}}\frac{\partial \xi_{kl}}{\partial u_{i,j}},
\end{equation}
there is an alternative way to compute the stress tensor for the linear case, i.e. neglecting $u_{k,i}u_{k,j}$ in the expression \eqref{strain}
\begin{equation}\label{linearstrainstress}
\sigma_{ij}=\frac{\partial \,U(\pmb{\xi})}{\partial\xi_{ij}},
\end{equation}
which is the usual form in linear deformation theory. For the nonlinear case there is an alternative way to compute the stress tensor as 
\begin{equation}\label{strainstress}
\sigma_{ij}=J_{ik}\frac{\partial \,U(\pmb{\xi})}{\partial\xi_{kj}},
\end{equation} 
where the summation convention is evoked and $\pmb{J}=\{J_{ij}\}_{ij}$ is the Jacobian matrix associated to the transformation from undeformed ($x_j$) to deformed ($x'_i$) coordinates,  whose elements are $J_{ij}=\partial_{x_j}x'_i$,  or  equivalently are $J_{ij}=u_{i,j}+\delta_{ij}$. For a complete demonstration see the Appendix \ref{ApendA}\label{AppA}. 
The stress tensor \eqref{strainstress} it is know as the \textit{first Piola-Kirchoff} stress tensor  \citep{Marsden,norris1998finite}.

The potential energy $U$, also called the elastic strain energy of the body, is an scalar quantity of the strain tensor $\pmb{\xi}$ and also for a given point of the body is invariant under spatial rotations 
for \textit{isotropic materials} \citep{murnaghanfinite}.  Therefore, $U$  depends on the invariants of the tensor $\pmb{\xi}$, given by: 
\begin{equation}\label{invariants}
I_n(\pmb{\gamma})=tr(\pmb{\gamma}^n).
\end{equation}
Note that a material is said to be \textit{isotropic} if it is \textit{unaffected by rotations}. This means that if the body is rotated and then do an experiment upon it, the outcome is the same as if the body had not been rotated, in other words, rotations cannot be detected by any experiment.

In general, for a $2-$rank and symmetric tensor, $\pmb{\gamma}$, which is represented as a matrix, there are 3 independent tensor invariants for example $(I_1, I_2, I_3)$. 

For isotropic materials, the general form of the elastic strain energy only depends on the invariants of $\pmb{\xi}$, e.g.  $I_n(\pmb{\xi})$ see expression \eqref{invariants}. Note that this set of invariants it's not the only one, there are others triplets of independent invariants for this kind of tensors \citep{murnaghanfinite}.  Under this hypothesis, the elastic stress tensor $\pmb{\sigma}$, given by \eqref{strainstress}, can be computed more easily using the equation
\begin{equation}
\partial_{\pmb{\xi}}I_n(\pmb{\xi})=n\;\pmb{\xi}^{n-1},
\end{equation}
for all $n\in \mathbb{N}$ \citep{lutkepohl}, this is one of the reasons to use this particular set of tensorial invariants. Using that and the chain rule,   the elastic stress tensor $\pmb{\sigma}$ is  
\begin{equation}\label{stressfuncenergy}
\pmb{\sigma} = \pmb{J}\sum_{n\geq 1}\; n\; \pmb{\xi}^{n-1}\; \partial_{I_n}U.
\end{equation}
The Cayley$-$Hamilton theorem \citep{Frobenius} applied to $\pmb{\xi}$ allow us to obtain its invariants: $I_n$ with $n\geq 4$, as a function of the previous and independents invariants $(I_1,I_2,I_3)$. For this reason,  in case to consider the expression \eqref{stressfuncenergy} to obtain $\pmb{\sigma}$,  
it is not necessary to include the $\xi-$invariants beyond order 3 in the potential  energy  $U$
\begin{equation}\label{funcenergy}
U(\pmb{\xi}):= U(I_1(\pmb{\xi}),I_2(\pmb{\xi}),I_3(\pmb{\xi})),
\end{equation}
where $U$ is (re)defined as a particular function of the first 3 of invariants of $\pmb{\xi}$, under the isotropic material hypothesis. 

In order to include a nonlinear behaviour in the dynamics equations, then $U $ must depend of the invariants of $\pmb{\xi} $ at order beyond 2. At third order in $\pmb{\xi}$, the potential energy is expressed from the invariants of $\pmb{\xi}$  as  \citep{landau86} 
\begin{equation}\label{U Landau}
U(\pmb{\xi}) = \tfrac{1}{2} \lambda I_1^2(\pmb{\xi})  +  \mu  I_2(\pmb{\xi}) + \tfrac{1}{3} \mathcal{A} I_3(\pmb{\xi}) + \mathcal{B} I_1(\pmb{\xi})I_2(\pmb{\xi}) +\tfrac{1}{3} \mathcal{C} I_1^3(\pmb{\xi}), 
\end{equation}
where $(\lambda,\mu)$ are the Lamé moduli and $(\mathcal{A,B,C})$ are the  Landau moduli.  At fourth order in $\pmb{\xi}$ the potential energy is expressed  as  \citep{hamilton2004nonlinear} 
\begin{equation}\label{U Hamilton}
\widetilde{U}(\pmb{\xi}) = U(\pmb{\xi}) +\mathcal{E}I_1(\pmb{\xi}) I_3(\pmb{\xi}) + \mathcal{F}I_1^2(\pmb{\xi}) I_2(\pmb{\xi})+\mathcal{G}I_2^2(\pmb{\xi})+\mathcal{H}I_1^4(\pmb{\xi}).
\end{equation}

Using the expressions \eqref{stressfuncenergy} and \eqref{U Hamilton} the elastic stress tensor   $\pmb{\sigma}$ is expressed as 
\begin{align}\label{elastic stress}
\pmb{\sigma}=&\,\pmb{J}\big[\lambda I_1(\pmb{\xi})+\mathcal{B}I_2(\pmb{\xi})+\mathcal{C}I_1^2(\pmb{\xi})+\mathcal{E}I_3(\pmb{\xi})+2\mathcal{F}I_1(\pmb{\xi})I_2(\pmb{\xi})+  4\mathcal{H}I_1^3(\pmb{\xi})\big]\nonumber\\
&\\
+&2\pmb{J\xi}\big[\mu+\mathcal{B}I_1(\pmb{\xi})+\mathcal{F}I_1^2(\pmb{\xi})+2\mathcal{G}I_2(\pmb{\xi}) \big]+3\pmb{J\xi}^2\big[\tfrac{1}{3}\mathcal{A}+\mathcal{E}I_1(\pmb{\xi})\big].\nonumber
\end{align}
In this work the nonlinear behaviour of the equation \eqref{dynamics} comes from two fundamental sources, namely \textit{intrinsic} and \textit{extrinsic}.  
The first of these is given by the nonlinear terms \textit{inside} the strain tensor $\pmb{\xi}$, see \eqref{strain}.  
In order to characterize the intrinsic source of nonlinearity, a bivalued parameter $\ell=0,1$ is introduced in $\pmb{\xi}$ as
\begin{equation}\label{strainn}	
\xi_{ij}=\frac{1}{2}\big(u_{i,j}+u_{j,i}+\ell \,u_{k,i}u_{k,j}\big).
\end{equation}
If the switch is \textit{off} (\textit{on}) there is a linear (nonlinear)  behaviour from \eqref{strainn}. In other words, $\ell=0$ ($\ell=1$) corresponds to small (large) strains given by \eqref{strainn}.

The second source of nonlinear behaviour comes to the tensor invariants of $\pmb{\xi}$ at order beyond $2$ inside the elastic strain energy $U$. A proper selection of the elastic strain energy involves a specific constitutive law and, consequently, a dynamical equation from the displacement field $\pmb{u}$. 

In the quasi-static deformation case, the elastic stress tensor $\pmb{\sigma}$ comes from a potential $U$, and given the latter as a function of the strain tensor, it was possible to write a relationship between elastic stress and deformation \eqref{elastic stress}. From this relationship, the nonlinear coefficients will be obtained explicitly in section \ref{Westervelt from Elasticity}, including $\beta$ of the Westervelt equation \eqref{WesterveltEq}, among others.

\subsection{Irreversible deformation process}\label{Irreversible process}

The aim of this subsection is to describe the irreversible deformation process in order to find the stress tensor associated with the dissipative effects, because there is an implicit relationship between  diffusivity, quantified by $\delta$ in the Westervelt equation \eqref{WesterveltEq}, with the loss of energy in the wave propagation.

In the previous section the body deformation was assumed as a quasi-static process as an approximation of a reversible process. So far, the local deformation of the material takes place in a way that the effect of its surrounding part is neglected. In reality, when a deformation is performed on a local region of the material its neighborhood exerts a resistance which tends to retard the motion. These delay effects increase as the deformation does the same, in particular, the more rapidly change it in time, which is the velocity of the deformation.      

Actually, the deformation process is thermodynamically irreversible, because the velocities involved in the deformation process are non-negligible, the body is not in equilibrium at every instant, but there is an internal processes that tend to return it to equilibrium. The irreversible character of the deformation process implies a loss of energy that is finally dissipated by being converted into heat. Essentially, there is two mechanism of dissipation of energy: \textit{viscosity} and \textit{thermal conduction}. At the molecular level, this two mechanism can be related \citep{Mohanty51,O'Neal62}, due to the non-negligible velocities involved in the deformation process. However, from the mechanical point of view there is no interest in the properties of the constituents of matter \textit{per se}, not even in the notion of the constituent of matter. Finally, the framework of deformation theory of the continuum media consists in the study of the mechanical state of the system, i.e. state of movement, with the addition of the thermodynamic state, due to certain collective properties of the continuous medium.

Regarding to viscosity, the delay effect on the propagation of the deformation can be understood as a force, which depends on the velocities of the deformation. The thermal conduction requires, on the other hand, a non-uniform distribution of temperatures on the material. The theoretical explanation of this mechanism can be derived from a \textit{heat continuity equation}, which is constructed from the following idea: the amount of heat absorbed per unit time in unit of volume of the body, $T\partial_tS$ (where $S$ is the entropy per unit volume), plus to the divergence of the heat flux density, from Fourier equation $\nabla\pmb{\cdot}(k \nabla\,T)$, must be equal to zero
\begin{equation}\label{Heat equation}
T\partial_tS+\nabla\pmb{\cdot}(k \nabla\,T)=0
\end{equation}
where $ k$ is the thermal conductivity of the body. The non-uniform distribution of the temperature is required and corresponds to an entropy production of the system, which that accounts for the irreversible character of the dissipative process. Even in the case in which the material can be monitored and controlled in order to maintain a uniform average temperature, this does not imply that there are local  fluctuations of temperature, that give rise to the effects of heat conduction, via the indicated mechanism. Finally, the required distribution of temperatures in a local region of the body implies changes of its volume in time: dilation and contraction. These effects increase as the more rapidly change the volume, i.e. the velocity of the thermal conduction depends on the velocity of the volumetric deformation.


For these reasons the viscous and thermal conduction effects must be described through dissipative forces,  therefore they should not be introduced from the elastic potential energy, as has been done in the previous section.
A consistent way of introducing these effects can be done through the formalism of the \textit{dissipative function}  \citep{landau86,landau80}.  Generally speaking, a mechanical system, whose motion involves a loss of energy can be understood from the \textit{dissipative forces} on the body. Nevertheless, taking on a certain spatial region inside the body,  this dissipative forces acts only on its  surface. Therefore this forces comes through another tensor, $\pmb{\sigma}'$, namely \textit{dissipative stress tensor}. In the last section, the elastic stress tensor was obtained from the elastic strain energy $U$, equations 
\eqref{first stress} 
or 
\eqref{strainstress}. In this section, the dissipative stress tensor is obtained from the dissipative function $\Psi$, that depends on partial time derivative of the  tensor $\pmb{\omega}$, where $\omega_{ij}=\frac{1}{2} (u_{i,j}+ u_{j,i})$, and  the most general form of a dissipative function is given by  \citep{landau86,landau80}
\begin{equation}\label{Psi}\Psi=\frac{1}{2}\,\eta_{ijkl}\,\partial_t \,\omega_{ij}\,\partial_t \,\omega_{kl}.\end{equation}
The connection between $\pmb{\sigma}'$ and $\Psi$ is given by
\begin{equation}\label{viscPsi}
\sigma'_{ij}=\frac{\partial\, \Psi}{\partial({\partial_t\omega_{ij}})},
\end{equation}
which is similar to expression \eqref{first stress}. 
Finally, from \eqref{viscPsi} and \eqref{Psi} 
\begin{equation}\label{stress viscoso}
\sigma'_{ij}=\eta_{ijkl}\,\partial_t\, \omega_{kl}.
\end{equation}
In order to give a concrete form of the tensor $\pmb{\eta}$, whose components $\{\eta_{ijkl}\}$,  the isotropic material hypothesis is evoked again. For this kind of materials there are two independent components of the tensor $\pmb{\eta}$. The dissipative stress tensor comes from an specific dissipative function $\Psi$ \citep{landau86,landau80} which depends on the first and second tensor invariants \eqref{invariants} of $\partial_t\, \pmb{\omega}$ 
\begin{equation}\label{dissipative function0}
\Psi=\eta\Big(\partial_t\,\omega_{ij}-\frac{1}{3} \delta_{ij}\partial_t\omega_{kk}\Big)^2 + \frac{1}{2}\big(\zeta+\chi\big)\big(\partial_t\,\omega_{kk}\big)^2 
\end{equation} 
where $\eta$ and $\zeta$, both positives, are denominated shear and bulk viscosity, respectively. The last parameter $\chi=k T \alpha_p^2/\rho \,c_p c_{_{V}}\alpha_{_{T}}$  \citep{norris1998finite}, corresponds to thermal conduction, defined through the thermal conduction $k$,  temperature $T$, density $\rho$, specific heat capacities at constant pressure and volume  $(c_p, c_{_{V}})$, and 
isobaric thermal expansion and  isothermal compressibility  $(\alpha_p,\alpha_{_{T}})$, defined by 
\begin{equation}
\alpha_p:=\frac{1}{V}\left( \frac{\partial V}{\partial T}\right)_p, \quad \alpha_{_{T}}:=-\frac{1}{V}\left(  \frac{\partial V}{\partial p}\right)_{_{T}},
\end{equation}
where the notation shows that the partial derivatives are computed keeping fixed the variables indicated as subscripts. 
Also, for small temperatures differences in the body that quantities such as, $k$, $c_p$, $c_{_{V}}$, $\alpha_p$, $\alpha_{_{T}}$, may be regarded as constants. It is possible to reduce the thermionic  coefficient $\chi$ using the generalized Mayer's relation $c_p-c_{_{V}}=T \alpha_p^2/\rho \alpha_{_T}$, then 
\begin{equation}
\label{chi, para los amigos}
\chi=k\left(\frac{1}{c_{_V}} - \frac{1}{c_p} \right).
\end{equation}

The dissipative function $\Psi$ can be expressed in terms of the tensor invariants of $\partial_t\,\pmb{\omega}$ as 
\begin{equation}\label{dissipative function}
\Psi=\eta\,I_2(\partial_t\,\pmb{\omega})+\left(\frac{\zeta+\chi}{2}-\frac{\,\eta\,}{3}\right) I_1^2(\partial_t\,\pmb{\omega})
\end{equation}
which is useful to compute the dissipative stress tensor using the equation \eqref{viscPsi} and the chain rule. For more details about the equivalent way to express $\Psi$ in \eqref{dissipative function0} and \eqref{dissipative function} see Appendix \ref{ApendB}.
Note that, given the indicated dissipative function $\Psi$ in order to define the dissipative stress tensor, there is no necessity to distinguish small or large deformations. In other words, the tensor $\pmb{\sigma}'$ depends on the partial time derivative of  $\pmb{\omega}$, which is the linear part of $\pmb{\xi}$, because the viscous effects depends on the relative velocity, given by time derivatives of the displacement field $\pmb{u}$ only \citep{landau86,landau80}.

The dissipative forces presents in the tensor $\pmb{\sigma}'$ must be added to the forces provided from the elastic stress tensor $\pmb{\sigma}$, introduced in section \ref{quasistatic process}, specifically in equation \eqref{first stress}. Using the expressions \eqref{viscPsi} and \eqref{dissipative function} the dissipative tensor   $\pmb{\sigma}'$ is expressed as
\begin{equation}\label{dissipative stress}
\pmb{\sigma}'=2\eta\,\partial_t\,\pmb{\omega}+\left(\zeta+\chi-\frac{\,2\eta\,}{3}\right)I_1(\partial_t\,\pmb{\omega})\pmb{I}
\end{equation}
where $\pmb{I}$ is the identity $3\times 3$ matrix.

The viscous coefficients $(\eta,\zeta)$ and the thermionic coefficient $\chi$ can be put in terms of the so called \textit{sound diffusivity} $\delta$, this will be the work of the next section. Once a differential equation of the Westervelt form is obtained, it will be observed that the multiplicative coefficient to the temporal derivative of third  order must be  precisely $\delta/c_l^2$, by comparison with \eqref{WesterveltEq}.

In the irreversible deformation case the dissipative stress tensor $\pmb{\sigma}'$ comes from a dissipative function $\Psi$, and given the latter as a function of the time derivative of the strain tensor, it was possible to write a relationship between dissipative stress and time derivative of deformation \eqref{dissipative stress}. From this relationship, the nonlinear coefficients will be obtained explicitly in section \ref{Westervelt from Elasticity}, including $\delta$ of the Westervelt equation \eqref{WesterveltEq}, among others.

In summary, considering a nonlinear behaviour and the mentioned dissipative effects, the $i-$component of involved forces per unit of undeformable volume \citep{landau86} are given, respectively, from the \textit{elastic} and \textit{viscous} stress tensors as
\begin{equation}\label{involved forces}
\partial_{x_j}\sigma_{ij}+\partial_{x_j}\sigma'_{ij}.
\end{equation}
For an exhaustive discussion about the dissipative function see \citep{landau86,landau87,landau80}.

\subsection{P$-$wave restriction and planar propagation}  \label{Pwaves}  

This works will be dealt with the propagation in one spatial dimension, for this reason the expressions for the elastic and the dissipative stress tensors  will be remarkably simplified. 
The one$-$dimensional nonlinear equation is a concrete case of study in order to show some results, which involves the study of a nonlinear partial differential equation, see \citep{polyanin2003handbook}. 
Two conditions that modifies the wave propagation are considered: $u_2 = 0 = u_3$, for all $(x_1,x_2,x_3)$  and $u_1$ depends on $x_1$ and time $t$ only. This two conditions corresponds to $P-waves$ restriction and \textit{planar} wave propagation, respectively. The strain tensor is reduced significantly under this two conditions: $\xi_{ij}=0, \forall\, i,j\neq 1$.  A simplified notation is taken which eliminate the spatial index, because there is only one,  namely replace $x$ instead $x_1$. Under the above hypothesis the displacement field and  the strain tensor are reduced to 
\begin{equation}\label{u xi}
\begin{matrix}
u_{i}(\pmb{x},t)\longmapsto  u(x,t)\delta_{i1},
&\xi_{ij}(\pmb{x},t)\longmapsto \xi(x,t)\delta_{i1} \delta_{j1}.\\
\end{matrix}
\end{equation}
A straightforward matrix representation of the strain tensor $\pmb{\xi}$ is $diag(\xi,0,0)$, similarly for $\pmb{\omega}$ is $diag(\partial_x u,0,0)$. Note that each natural power of a diagonal matrix is also another diagonal matrix, then $I_n(\pmb{\xi})=\xi^n$ and $I_n(\partial_t\,\pmb{\omega})=[\partial_t\,(\partial_xu)]^n$.

It has been said that the elastic stress tensor $\pmb{\sigma}$  and the dissipative stress tensor $\pmb{\sigma}'$ are calculated from the potential energy $U$ and dissipative function $\Psi$, which are functions of the tensor invariants of $\pmb{\xi}$ and $\partial_t\,\pmb{\omega}$, see \eqref{elastic stress} and \eqref{dissipative stress} respectively, both stress tensors are reduced to a diagonal form, but even in this case ($P-$waves and planar propagation), the components 
$(\sigma_{22},\sigma_{33})$ and $(\sigma'_{22},\sigma'_{33})$ are not necessary equal to zero. For an explicit expression of each components of the stress tensors see Appendix \ref{ApendC}.



The purpose of this section is obtain a differential equation for the displacement field, under the $P-$wave restriction and planar propagation. Is more convenient the notation \eqref{u xi} for $\pmb{u}$ and $\pmb{\xi}$. The non-zero forces per unit volume given by \eqref{involved forces} are reduced to $\partial_x\sigma_{11}+\partial_x\sigma'_{11}$, then $\sigma_{11}$ and $\sigma_{11}'$ are simply denoted by $\sigma$ and $\sigma'$.

In the \textit{quasi-static regime} a possibility of large deformations in the strain tensor \eqref{strainn} is considered. Therefore the fundamental expressions in deformation theory corresponds to the kinematics, the dynamical equilibrium and a nonlinear constitutive law, are reduced to 
\begin{align}\label{new dynamics}
\begin{aligned}
\xi(x,t)&=\partial_x u(x,t)+\tfrac{1}{2}\,\ell\,[\partial_x u(x,t)]^2\\
0&=\rho \,\partial_{tt}u(x,t)-\partial_x\sigma(x,t)\\
\sigma&=F(\xi)\quad or\quad \sigma=G(\partial_x u).
\end{aligned}
\end{align}


From the elastic strain energy at fourth order in $\pmb{\xi}$ \eqref{U Hamilton} 
\begin{equation}\label{1Denergy}
U(\xi) = \mathlarger{\tfrac{1}{2}} a\,\xi^2 + \mathlarger{\tfrac{1}{3}} b\, \xi^3+\mathlarger{\tfrac{1}{4}} c\, \xi^4.
\end{equation}

In order to obtain a differential equations in terms of displacements, e.g. $\sigma=G(\partial_x u)$ is given by
\begin{equation}\label{SIGMA}
\sigma=a_1\,\partial_xu+a_2(\partial_xu)^2+a_3(\partial_xu)^3+ \mathcal{O}\pmb{(}(\partial_x u)^4\pmb{)},
\end{equation}
where the coefficient $(a_1,a_2,a_3)$ are given by
\begin{align}
a_1&=\lambda+2\mu\nonumber\\
a_2&=\frac{\;3\;}{2}a_1\ell+\mathcal{A+3B+C}\\
a_3&=\Big(\mathlarger{\tfrac{1}{2}}a_1+2a_2\Big)\ell+\mathcal{2E+4F+4G+4H}\nonumber
\end{align}
and  $\mathcal{O}$ is the \textit{order of function}, in Bachmann-Landau notation \citep{Bachmann,LandauE}. Finally $(\lambda,\mu)$, $(\mathcal{A,B,C})$ and $(\mathcal{E,F,G,H})$ are the Lam\'e, Landau and Hamilton parameters,  respectively, see expression \eqref{U Hamilton}. 


About the parametrization of $\sigma$, following others authors in \citep{hamilton1998nonlinear}, the coefficients $(a_1,a_2,a_3)$ in \eqref{SIGMA} can be reorganized and renamed as
\begin{equation}\label{SIGMA2}
\sigma=a_1\left[\partial_xu-\beta(\partial_xu)^2+\gamma(\partial_xu)^3+\mathcal{O}\pmb{(}(\partial_x u)^4\pmb{)}\right],
\end{equation}
where $\beta=-a_2/a_1$ and $\gamma=a_3/a_1 $. The parameters $(\beta,\gamma)$ can be expressed in terms of the LamÃ©, Landau, and Hamilton parameters as
\begin{align}\label{beta gamma}
\begin{aligned}
\beta &=-\frac{\;3\;}{2}\ell-\frac{\mathcal{A}+3\mathcal{B}+\mathcal{C}}{\lambda +2\mu}, \\[10pt]
\gamma&=\frac{\;\ell\;}{2}+2\frac{\ell(\mathcal{A}+3\mathcal{B}+\mathcal{C})+\mathcal{E+2F+2G+2H}}{\lambda +2\mu}.
\end{aligned}
\end{align}

In order to obtain a closed expression for $\partial_x\sigma$  in  \eqref{new dynamics}, the chain rule will be used in \eqref{SIGMA} as $\partial_x \sigma=\partial_{xx} u\,.\,\partial_{u'_x} \sigma $, where $\partial_xu \equiv u'_x$. Then the equation \eqref{new dynamics} becomes 
\begin{equation}\label{final dynamics}
\partial_{tt}u=c_l^2\partial_{xx}u\left[1+\alpha_1\partial_x u+\alpha_2(\partial_x u)^2\right],
\end{equation}
where the involves constants are $c_l=\sqrt{(\lambda+2\mu)/\rho}$  the longitudinal wave velocity, $\alpha_1=-2\beta$ and $\alpha_2=3\gamma$.  \label{comment} 

There is a nonlinear behaviour even in the \textit{small strains} regime, i.e. $\ell=0$, because the coefficient $\alpha_1\neq 0$ in this case. On the other hand, in the \textit{hookean} regime, i.e. $b=0$ and $c=0$ in the elastic strain energy \eqref{1Denergy}, there is a possibility of nonlinear behaviour from $\ell=1$, but is only geometrical. In this case the coefficient $\alpha_1=3$ and $\alpha_2=3/2$ are mathematical constants, there is no specific physical scales that characterize this nonlinear behaviour source once within this regime.

Given that the equation \eqref{final dynamics} contains terms of $\partial_xu$ at second-order, its solution can be founded from a perturbative method at first and second order. A solution of \eqref{final dynamics} can be expressed as 
\begin{equation}
u=\mathtt{u_0+u_1+u_2}
\end{equation}
The main idea of this method is to use a general solution, $\mathtt{u_0}$, of the zero$-$order problem, which in this case corresponds to $ \alpha_1=0=\alpha_2$ in \eqref{final dynamics}:
\begin{equation}\label{zeroorder}
\partial_{tt}\mathtt{u_0}=c_l^2\partial_{xx} \mathtt{u_0},
\end{equation}
and construct whit it the first$-$order term and the second$-$order term:  $\mathtt{u_1}$ and $\mathtt{u_2}$, respectively. 
The amplitudes of $\mathtt{u_1}$ and $\mathtt{u_2}$ can be put in terms of the parameters $\alpha_1$ and $\alpha_2$. A review about this method in the next subsection is presented. For more details see \citep{Bochud}.

A brief comment about the dynamics, there is a trivial generalization of the dynamical equation \eqref{final dynamics}. In general, $\sigma$ is expressed as a polynomial function of $\partial_x u$ at a certain order $n$ 
\begin{equation}\label{sigma generico}
\sigma=\sum_{k=1}^n a_k\,(\partial_xu)^k,
\end{equation}
where the chain rule will be used again as $\partial_x \sigma=\partial_{xx} u\,.\,\partial_{u'_x} \sigma $ and $\partial_xu \equiv u'_x$. Then the expression \eqref{new dynamics} becomes

\begin{equation}\label{final dynamics gen}
\partial_{tt}u-c_l^2\partial_{xx}u= c_l^2\partial_{xx}u\,\Bigg[ \sum_{k=2}^{n} \mathlarger{\frac{k\,a_k}{a_1}}(\partial_x u)^{k-1}\Bigg],
\end{equation}
where $c_l^2=a_1/\rho$. This is the general equation for $u(x,t)$ with nonlinear effects in quasi-static regime.

In the \textit{irreversible regime}, the dissipative stress tensor must be incorporated from \eqref{stress viscoso}, which is reduced to 
\begin{equation}\label{stress viscoso2}
\sigma'=\nu\,\partial_t (\partial_x u)
\end{equation}
where $\nu=4\eta/3+\zeta+ \chi$, $(\eta,\zeta)$ are the shear and bulk viscosity respectively and $\chi$, given by \eqref{chi, para los amigos}, characterize the thermal conduction effects. Then the differential equation for the displacement is given by
\begin{equation}\label{final dynamics gen viscosa}
\partial_{tt}u-c_l^2\partial_{xx}u= c_l^2\partial_{xx}u\,\Bigg[ \sum_{k=2}^{n} \frac{k\,a_k}{a_1}(\partial_x u)^{k-1}\Bigg]+\frac{\;\nu\;}{\rho}\, \partial_{txx}u.
\end{equation}
The solution of \eqref{final dynamics gen} and \eqref{final dynamics gen viscosa} can be founded through a perturbative method at a precision order  $\mathtt{N}$ as 
\begin{equation}
u=\sum_{\mathtt{k=0}}^\mathtt{N}\,\mathtt{u_k}.
\end{equation}
where $\mathtt{u_0}$ is a solution of the linear wave equation \eqref{zeroorder} and each of the next functions $\mathtt{u_k}$ is the $\mathtt{k}-$harmonics the fundamental solution $\mathtt{u_0}$.

\subsubsection{Non$-$linear behaviour at first order in quasi-static regime}
The case of the nonlinear wave equation up to the first$-$order nonlinearity is considered initially, corresponds to $\mathcal{O}\pmb{(}(\partial_x u)^2\pmb{)}\sim  0 $ in \eqref{final dynamics}, therefore, this equation is reduced to,
\begin{equation}
\partial_{tt}u=c_l^2\partial_{xx}u\,\big(1+\alpha_1\partial_x u\big).
\end{equation}
Following the perturbation theory \citep{hamilton1998nonlinear} allows to decompose the wave displacement at first-order as,
\begin{equation}
u=\mathtt{u_0+u_1}.
\end{equation}
Then the solution for $\mathtt{u_1}$ is coming from $\mathtt{u_0}$, in \eqref{zeroorder}, as 
\begin{equation}\label{firstorder}
\partial_{tt}\mathtt{u_1}-c_l^2\partial_{xx}\mathtt{u_1}=c_l^2\alpha_1\,\partial_{xx} \mathtt{u_0}\,\partial_{x}\mathtt{u_0}.
\end{equation}
The parameter $\alpha_1$, or $-2\beta$, in the other parametrization \eqref{SIGMA2} and \eqref{beta gamma}, can be determined from the $\mathtt{u_1}$ amplitude  \citep{Bochud}.

\subsubsection{Non$-$linear behaviour at second order in  quasi-static regime}

The case of the nonlinear wave equation up to the second$-$order nonlinearity is considered now, and corresponds to $\mathcal{O}\pmb{(}(\partial_x u)^3\pmb{)}\sim  0 $ in \eqref{final dynamics}, therefore, this equation is
\begin{equation}
\partial_{tt}u=c_l^2\partial_{xx}u\left[1+\alpha_1\partial_x u+\alpha_2(\partial_x u)^2\right].
\end{equation}
Now, from the perturbation theory \citep{hamilton1998nonlinear} allows to decompose the wave displacement at second-order as,
\begin{equation}
u=\mathtt{u_0+u_1+u_2}.
\end{equation}
Then the solution for $\mathtt{u_2}$ is coming from $\mathtt{u_0}$ \eqref{zeroorder} and $\mathtt{u_1}$ \eqref{firstorder}, as 
\begin{equation}\label{secondorder}
\partial_{tt}\mathtt{u_2}-c_l^2\partial_{xx}\mathtt{u_2}=c_l^2 \alpha_1\,\partial_{x}\left(\partial_x \mathtt{u_0}\partial_x \mathtt{u_1}\right)+c_l^2 \alpha_2\,\partial_{xx} \mathtt{u_0}\,(\partial_{x}\mathtt{u_0})^2.
\end{equation}
The parameter $\alpha_2$, or $3\gamma$, in the other parametrization \eqref{SIGMA2} and \eqref{beta gamma},  can be determined from the $\mathtt{u_2}$ amplitude  \citep{Bochud}.

\section{Derivation of a generalized Westervelt equation}\label{Westervelt from Elasticity}

In this section the equations of the body deformations will be related to the propagation of sound in this media. This propagation will be manifested  in terms of pressure waves, in order to obtain the dynamics proposed by Westervelt equation.  

The bridge between the stress tensors in deformation theory and pressure comes from the definition of the complete stress tensor, $\pmb{\sigma}+\pmb{\sigma}'$, including the quasi-static and the irreversible processes. Each component of this complete tensor $\sigma_{ij}+\sigma_{ij}'$ represents the $i-$component of the force on unit area perpendicular to the $x_j-$axis. Therefore the $i-$component force on the surface perpendicular to $x_i-$axis, namely $p_i$, is related by the complete stress tensor as: $p_i=-(\sigma_{ii}+\sigma_{ii}')$, see section \ref{signo fuerzas}. In other words, the diagonal elements of the complete stress tensor contains the forces per unit of area, with no summation convention \citep{landau86}.

Regarding the concept of \textit{hydrostatic pressure}, $p$, in fluid dynamics context is defined from the complete stress tensor, $\pmb{\Sigma}$, as 
\begin{equation}
p:=-\frac{1}{3}I_1(\pmb{\Sigma}),
\end{equation}
this definitions has a underlying assumption: $\Sigma_{11}=\Sigma_{22}=\Sigma_{33}$, which is valid in fluid dynamics. In other words $p$ can be calculated from the \textit{arithmetic mean} (or in general, from a uniformly pondered averaged value) of $\{\Sigma_{11},\Sigma_{22},\Sigma_{33}\}$. This assumption it can recovered exactly in linear deformation theory  neglecting the shear forces, which corresponds to quasi-fluids. The elastic stress tensor is obtained from  $U(\pmb{\xi}) = \tfrac{1}{2} \lambda I_1^2(\pmb{\xi})  +  \mu  I_2(\pmb{\xi})$ as a particular case of the expression \eqref{U Landau}, then $\pmb{\sigma}=\lambda I_1(\pmb{\xi})\pmb{I}+2\mu\,\pmb{\xi}$ obtained from the equation \eqref{strainstress}.  A sufficient condition that guarantees  $\sigma_{11}=\sigma_{22}=\sigma_{33}$ corresponds to take $\mu=0$. However the last assumption it's not exactly valid in the nonlinear deformation theory applied to solids or quasi-fluids. In the nonlinear case the condition $\mu=0$ does not implies $\sigma_{11}=\sigma_{22}=\sigma_{33}$. Nevertheless, for the case of $P-$waves and planar propagation $\sigma_{22}=\sigma_{33}$ and the difference between $\sigma_{11}$ and $\sigma_{22}$ is small, 
because the potential energy $U$ depends on the tensor invariants beyond order $2$. Moreover the dissipative forces presents in $\pmb{\sigma}'$ \eqref{dissipative stress} satisfies that $\sigma'_{11}=\sigma'_{22}=\sigma'_{33}$ under $\eta=0$. This implication is also valid in nonlinear deformation theory. 

In section \ref{Pwaves} it has been said that for $P-$waves  and planar propagation the non-zero forces per unit volume are given by $\partial_x\sigma+\partial_x\sigma'$, thus $\sigma_{11}=\sigma$ and $\sigma_{11}'=\sigma'$ are the relevant components of the complete stress tensor. 
\begin{equation}\label{p-sigma}
p=-\sigma-\sigma'.
\end{equation}
this is the \textit{solid pressure}, i.e. the force per unit area,
with a region exerts on the surface surrounding in the $x-$direction, see definition in section \ref{nonlinear continuous media}. For a detailed complement see the \ref{ApendC}.

The main idea for the deduction of a partial differential equation for $p$ consist in replace the expression \eqref{p-sigma} into the dynamical equation of the deformation theory  $\rho\partial_{tt} u = \partial_x \sigma+\partial_x \sigma'$ \eqref{new dynamics} and re-write it as
\begin{equation}\label{generator}
\rho \partial_{tt}(\partial_x u)=-\partial_{xx}p
\end{equation}  
in order to obtain a second order spatial-derivative in $p$. Thereby, the central problem of the following two sections corresponds to obtain an inverted expression of $\partial_x u$ in terms of the pressure $p$, and eventually $\partial_t p$,  with the propose to obtain the Westervelt equations an its generalization.

A comment about the nature of the terms of the Westervelt equation. It has been said that the Westervelt equation \eqref{WesterveltEq} is written so that the left member is associated with linear waves, which can be obtained form a linear and reversible theory of deformations. On the other hand, the right member is composed by two terms, related to specific effects of waves propagation: nonlinearity and diffusivity. 
This two effects in the Westervelt equation will be reconstructed from a nonlinear and irreversible theory of deformations. Firstly, the nonlinear term in $p$,  $\partial_{tt} (p^2)$, quantified by $\beta$ comes from a nonlinear strain energy $U$, proposed by Landau \citep{landau86},  which defines a nonlinear elastic stress tensor $\sigma$, given by \eqref{sigma generico}. 
Finally, the sound diffusivity term in $p$, $\partial_{ttt} (p)$, quantified by $\delta$ corresponds to an irreversible process. Generally speaking, the irreversibility must be manifested from a symmetry breaking under temporal inversion of the dynamics.

In order to identify the effects of these two terms of equation  \eqref{WesterveltEq} from the deformation theory, a deduction in two stages is proposed: \textit{quasi-static} and \textit{irreversible} regimes in the next sections, within the same $P-$waves and plane waves context discussed in the section \ref{Pwaves}.

\subsection{Quasi-static regime}\label{NON VISCOUS}

The deformation in the elastic bodies is assumed reversible. It has been said  that all the forces come from a potential energy $U$, described in subsection \ref{Pwaves}. Generally speaking, a material wave propagates as a localized pressure change. Therefore, a dynamics that govern these pressure changes must be funded, a differential equation for $p$ can will be obtained in this section. 

From expressions \eqref{sigma generico} and \eqref{p-sigma} the pressure $p$ is  
\begin{equation}\label{Pressure}
p= \OP{P}(\partial_x u),
\end{equation}
where $\OP{P}$ is a polynomial function, which is equal to $-\sigma$
\begin{equation}\label{P}
\OP{P}(\partial_xu)=-\sum_{l=1}^n a_l (\partial_xu)^l.
\end{equation}
where $a_1\neq 0$ is considered, which is guaranteed because $a_1=\lambda +2\mu\neq 0$, see equation \eqref{SIGMA}.

Consider now  an inverse function of $\OP{P}$, denoted by $\OQ{Q}$
\begin{equation}\label{inversePressure}
\partial_x u= \OQ{Q}(p).
\end{equation}
Finally, replacing  the expression \eqref{inversePressure} into the dynamical equation \eqref{generator} can be obtain 
\begin{equation}\label{WesterbeltGeneralizator}
\rho \partial_{tt}\OQ{Q}(p)=-\partial_{xx}\,p.
\end{equation}
which is a differential equation for $p(x,t)$. 
 The diversity of situations that can be described by \eqref{WesterbeltGeneralizator}, i.e. its \textit{epistemological robustness}, comes from the variety of possible polynomials $\OQ{Q}(p)$. In addition, this variety of $\OQ{Q}$ is limited  by $\OP{P}$,  because $\OQ{Q}$ is the inverse function of $\OP{P}$. 

The main problem of this section involves building an expression to obtain $\OQ{Q}$ from $\OP{P}$.  Regarding the calculation of $\OQ{Q}$ without loss of generality it can be expressed as Taylor series 
\begin{equation}\label{Q}
\OQ{Q}(p)=\sum_{k\geq 1} b_k\, p^k
\end{equation}
given that $\OQ{Q}$ is obtained through the inversion of $\OP{P}$, its respective coefficients $\{a_l\}_l$ and $\{b_k\}_k$,  are strongly related. The first four coefficients  $\{b_1,b_2,b_3,b_4\}$ for $\OQ{Q}$ are listed below
\begin{equation}\label{coefQ}
\begin{matrix}
b_1=-\dfrac{\;1\;}{a_1} &&b_2=-\dfrac{\;a_2\;}{a_1^3}\\
&\\
b_3=-\dfrac{\;2a_2^2-a_1a_3\;}{a_1^5}& &b_4=\dfrac{\;5a_1a_2a_3-a_1^2a_4-5a_2^3\;}{a_1^7}
\end{matrix}
\end{equation}

In particular, the coefficients $\{b_1,b_2,b_3,b_4\}$ from \eqref{coefQ} can be obtained using the specific coefficients $\{a_1,a_2,a_3,a_4\}$ from \eqref{SIGMA2}
\begin{equation}\label{b1,b2,b3,b4}
\begin{matrix}
b_1=-(\lambda+2\mu)^{-1} & &b_2=\beta(\lambda+2\mu)^{-2}\\
&\\
b_3=(\gamma-2\beta^2)(\lambda+2\mu)^{-3}& &b_4=5(\beta^3-\beta\gamma)(\lambda+2\mu)^{-4}.
\end{matrix}
\end{equation}

Following  \citep{Morse,Abramowitz}, the general structure for each coefficient $b_k$, $k\in\mathbb{N}$ is:
\begin{equation}
b_k=\frac{(-1)^k}{k!}\frac{d^{k-1}}{ds^{k-1}}\left(\sum_{l=1}^{n}a_{l}s^{l-1}\right)^{-k}\Bigg\vert_{s=0}.
\end{equation}

Replacing $\OQ{Q}(p)$ from \eqref{Q} in \eqref{WesterbeltGeneralizator}, an  explicit equation for $p(x,t)$ in the quasi-static regime is obtained as:
\begin{equation}\label{eqgenericP}
\sum_{n \geq 1}\rho\, b_n \partial_{tt}\pmb{(}p^n(x,t)\pmb{)}+\partial_{xx}\,p(x,t)=0,
\end{equation}
This expression allow to express the generalization of Westervelt equation in this regime for high order terms in pressure. Is possible to express this equation in a way that it  appears the linear wave part and the nonlinear terms separately, a simple way to do that is dividing both members by $\rho b_1$ (equal to $-c_l^{-2}$)
\begin{equation}\label{eqgenericP2}
\partial_{tt}\,p(x,t)-c_l^2\partial_{xx}\,p(x,t)=-\sum_{n \geq 2} \, b_nb_1^{-1} \partial_{tt}\pmb{(}p^n(x,t)\pmb{)}.
\end{equation}
The left hand correspond to the linear wave and the right hand contains the nonlinear terms as an inhomogeneous part. Note that both equations for $u(x,t)$ \eqref{final dynamics} and  $p(x,t)$  \eqref{eqgenericP2} are similar with respect to this general structure. The equation  for $p$ contains \textit{time derivatives} and the equation for $u$ contains \textit{spatial derivatives}, as an inhomogeneity terms. 



The Westervelt equation without dissipative effects can be recovered retaining the first term  in the sum of the differential equation \eqref{eqgenericP2} take the form 
\begin{equation}
\label{cuasiWestervelt}
\frac{\partial^2 p}{\partial t^2}-c_l^2\frac{\partial^2p}{\partial x^2}=\frac{\beta }{\rho c_l^2} \frac{\partial^2 p^2}{\partial t^2}
\end{equation}
where the nonlinear parameter $\beta$  is given by  \eqref{beta gamma}. 

This complete the deduction for the Westervelt equation  \eqref{WesterveltEq} from deformation theory and its generalization at any harmonic order in lossless energy regime.



\subsection{Irreversible regime}

In this context the energy is dissipated, ultimately in the form of heat. In $P-$waves and planar propagation, the stress tensor due to viscous and thermal conduction effects, $\pmb{\sigma}'$, in equation \eqref{stress viscoso} is reduced to \eqref{stress viscoso2}: $\sigma'=\nu\,\partial_t (\partial_x u)$, where $\nu=4\eta/3+\zeta+\chi$ and $(\eta,\zeta)$ are the shear and bulk viscosity respectively and $\chi$  from \eqref{chi, para los amigos}, characterize the thermal conduction effects.

It has been said that the there are two regimes: quasi-static and irreversible. The first, comes from the elastic stress tensor $\pmb\sigma$, which is expressed as a polynomial in $\partial_xu$, see equation \eqref{sigma generico} in section \ref{Pwaves}. The second, comes from the dissipative stress tensor $\pmb{\sigma}'$, given by \eqref{stress viscoso2}. Now, the connection between pressure and stress tensor is now $p=-(\sigma+\sigma')$, then 

\begin{equation}\label{p viscosa}
p=\OP{P}(\partial_x u)-\nu\;\partial_t (\partial_x u)
\end{equation}

In the previous section, the problem was solved considering an inverse polynomial of $\OP{P}$ namely $\OQ{Q}$, because in that case the time changes of $\partial_x u$ was neglected. In this section these effects have to be considered.  

In the introduction of section \ref{Westervelt from Elasticity} it has been announced about the necessity to obtain an inverted expression of $\partial_x u$  in terms of the pressure $p$ and $\partial_t p$, in order to obtain the Westervelt equation, now with the addition of loss energy term. Applying $\OQ{Q}$ in each member of \eqref{p viscosa}
\begin{equation}\label{p viscosa2}
\OQ{Q}(p)=\OQ{Q}\pmb{\big(} \OP{P}(\partial_x u)-\nu\;\partial_t (\partial_x u) \pmb{\big)}
\end{equation}
In the right member of \eqref{p viscosa}, its second term is smaller than its first term, comparatively speaking. This allows to express the right member of equation \eqref{p viscosa2} as 
\begin{align}\label{p viscosa3}
\OQ{Q}(p)&=\OQ{Q}\pmb{\big(} \OP{P}(\partial_x u)\pmb{\big)}-\OQ{Q}'\pmb{\big(}\OP{P}(\partial_x u)\pmb{\big)} \,\nu\;\partial_t (\partial_x u)
\end{align}
where $\OQ{Q}(x+y)=\OQ{Q}(x)+\OQ{Q}'(x)y+\cdots$ is used. Finally the first term of the left hand of \eqref{p viscosa3} is exactly $\partial_x u$, because $\OQ{Q}$ is the inverse polynomial of $\OP{P}$ and the second term of \eqref{p viscosa3} is written as 
\begin{equation}\label{p viscosa4}
\OQ{Q}(p)=\partial_x u-\nu\, b_1a\partial_t (\partial_x u)
\end{equation}
because each term $b_n\nu$, for $n>1$, is neglected, or is comparatibly smaller than $b_1\nu$. From expression \eqref{p viscosa4} at first order in this new perturbation parameter $ b_1\nu$, can be obtain $\partial_x u=\OQ{Q}(p) $. Taking the time derivative of the last expression: $\partial_t(\partial_x u)=\partial_t\OQ{Q}(p)$ and replace it into the expression \eqref{p viscosa4} \begin{equation}\label{generador}
\partial_x u=\OQ{Q}(p)+b_1\nu \partial_t\OQ{Q}(p)
\end{equation} Finally, replace the expression  \eqref{generador} into the dynamical equation \eqref{generator}
 \begin{equation}\label{WGenerator}
\rho\partial_{tt}\pmb{(}\OQ{Q}(p)\pmb{)}+\rho\,b_1\,\nu\, \partial_{ttt}\pmb{(}\OQ{Q}(p)\pmb{)}=-\partial_{xx} p.
\end{equation}
which is a differential equation for $p(x,t)$, similar to  \eqref{WesterbeltGeneralizator}, in fact, both differential equations are the same if $\nu=0$. Again, the diversity of situations that can be described by \eqref{WGenerator}, comes from the variety of possible polynomials $\OQ{Q}(p)$.

Replacing $\OQ{Q}(p)$ from \eqref{Q} in  \eqref{WGenerator}, an explicit  equation for $p(x,t)$ is obtained:
\begin{equation}\label{eqgenericP viscosa}
\sum_{n \geq 1}\rho\, b_n \partial_{tt}\pmb{(}p^n\pmb{)}+\rho\,b_1b_n\,\nu\;\partial_{ttt}\pmb{(}p^n\pmb{)}+\partial_{xx}p=0,
\end{equation}
Is possible to express this equation in a way that it  appears the linear wave equations and the nonlinear terms separately, again, a simple way to do that is dividing both members by $\rho b_1$ (equal to $-c_l^{-2}$)
\begin{equation}\label{eqgenericP2 viscosa}
\partial_{tt}\,p-c_l^2\partial_{xx}p=-\sum_{n \geq 2}  b_n b_1^{-1} \partial_{tt}\pmb{(}p^n\pmb{)}-\sum_{n \geq 1}\nu\, b_n\partial_{ttt}\pmb{(}p^n\pmb{)}.
\end{equation}
The left hand correspond to the linear wave and the right hand contains the nonlinear terms as an inhomogeneous part. Note that both equations for $u(x,t)$ \eqref{final dynamics} and  $p(x,t)$ \eqref{eqgenericP2 viscosa} are similar with respect to that general structure. The equation  for $p$ contains \textit{time derivatives} and the equation for $u$ contains \textit{spatial derivatives}, as an inhomogeneity terms. 

The generalized Westervelt equation \eqref{eqgenericP2 viscosa} presents a nonlinear behaviour even in the \textit{small strains} regime, i.e. $\ell=0$, because $\beta\neq 0$ and $\gamma=0$ and finally $b_2\neq 0$ and $b_3\neq 0$. On the other hand, in the \textit{hookean} regime, there is a nonlinear behaviour from $\ell=1$, but is only geometrical. In this case the coefficient $\beta=-3/2$ and $\gamma=1/2$ are mathematical constants, there is no specific physical scales that characterize this nonlinear behaviour source once within this regime.

The Westervelt equation with dissipative effects can be recovered as an approximation at non trivial first order of each sum of the differential equation \eqref{eqgenericP2 viscosa} 
\begin{equation}\label{Westervelt}
\frac{\partial^2 p}{\partial t^2}-c_l^2\frac{\partial^2p}{\partial x^2}=\frac{\beta }{\rho c_l^2} \frac{\partial^2 p^2}{\partial t^2}+\frac{\nu}{\rho c_l^2} \frac{\partial^3 p}{\partial t^3}
\end{equation}
where, by comparison with the Westervelt equation \eqref{WesterveltEq}, the parameters $\beta$ and $\delta=\nu/\rho$ are  given by
\begin{align}\label{beta delta}
\beta =-\frac{\;3\;}{2}\ell-\frac{\mathcal{A}+3\mathcal{B}+\mathcal{C}}{\lambda +2\mu}, \qquad
\delta=\frac{1}{\;\rho\;}\Big(\frac{\;4\;}{3}\eta+\zeta\Big)+\frac{k}{\;\rho\;}\Big(\frac{1}{c_{_{V}}}-\frac{1}{c_p}\Big).
\end{align}
Note that the parameter $\delta$ here in \eqref{beta delta} is exactly the same that appears in \eqref{deltadeWestervelt}, because the dissipative stress tensor was implemented here in the same way that in fluid dynamics \citep{landau86,landau87}. 

This complete the deduction for the Westervelt equation  \eqref{WesterveltEq} from deformation theory and with a general constitutive equation \eqref{sigma generico} with loss energy effects.

\section{Conclusions}

The Westervelt equation was obtained in fluid dynamics from a state equation $P=\varphi(\varrho,S)$, a continuity equation for the density, the Navier-Stokes equation for the velocity field, which includes the viscous forces, and finally the Fourier equation that describe the heat-transfer effects.


In this work the Westervelt equation \eqref{WesterveltEq} was written so that the left member are identified with linear waves and is obtained form a linear and reversible theory of deformations. On the other hand, the right member is composed by two terms, associated to specific effects of sound propagation: nonlinearity and diffusivity. Each of the last terms is quantified by $ \beta$ and $ \delta $, respectively. The last two effects in the Westervelt equation was reconstructed from a nonlinear and irreversible theory of deformations. The nonlinear term in $p$, i.e. $\partial^2_t (p^2)$ comes from a nonlinear strain energy $U$, proposed by Landau,  which defines a nonlinear elastic stress tensor. The concept of sound diffusivity  corresponds to an irreversible process, which is correctly implemented from $\partial_t^{3}p$ in Westervelt equation. In the context of the deformation theory, the diffusivity was related to \textit{irreversible} regime. In order to identify the effects of these two terms of Westervelt equation from the deformation theory, a successful deduction in two stages was proposed: \textit{quasi-static} and \textit{irreversible} regimes, within the same $P-$waves and plane waves context.

The role of small and large displacements are clearly important if the nonlinear effects on harmonic generation is taken into account. There are essentially two sources of nonlinear effects: \textit{intrinsic} or \textit{geometrical} and \textit{extrinsic} or \textit{physical}, that are clearly identified through the switch variable $\ell=0,1,$ for small and large deformations, and from the concrete elastic strain energy $U$, that depends on the invariants of the strain tensor at order beyond $2$. However, the inclusion of nonlinear viscous terms is still open. Likewise, the mathematical machinery developed in this work shows that this must be done through  a generalization to dissipative functions $\Psi$ that depends on tensor invariants of order beyond $2$.


It has been said that the Westervelt equation is implemented in the field of medical ultrasound for tissues and fluids, to model the high-intensity focused ultrasound propagation generated by medical ultrasound transducers. On the other hand, there is another way to describe nonlinear waves propagation through the so called  Khokhlov$-$Zabolotskaya$-$Kuznetsov equation \citep{Kuznetsov71,zabolotskaya69}, or KZK for short, which is valid for directional waves and can be applied for transducers with relatively small aperture angles. The KZK equation can be deduced from the Westervelt equation, under a change of variables \citep{hamilton1998nonlinear}. The Westervelt equation can be used also for large aperture-angle transducer. In terms of simulations there is an unique global solution for sufficiently small and regular data, which converges at an exponential rate as time tends to infinity \citep{Meyer2011}. There is local and global well-posedness as well as exponential decay for the Westervelt equation with inhomogeneous Dirichlet/Neumann boundary conditions \citep{Kaltenbacher2011}.

In this paper was considered $P-$waves propagation, the components of the displacement field $\pmb{u}$ they were considered as: $u_1$ is not identically zero, $u_2 = 0$ and  $u_3 = 0$. It is possible to change the propagation mode, modifying the condition on the displacement field in order to include another type of perturbations in the media, beyond $P-$wave propagation. In particular for $S-$waves,  $u_1=0$, $u_2$ and $u_3$ are not identically zero. There is  solid evidence that this kind of waves are more sensible to measure mechanical properties in soft tissue  \citep{GIAMMARINARO2018236}.


The deduction of Westervelt equation and even the proposed generalization within the context of deformation theory  allow us to consider future applications in tissue wave propagation, correlating these nonlinear parameters with tissue characterization or even with different pathologies in tissue. The simulation of these effects is useful for the correct assessment of the ultrasound dose required for the medical treatment. Combined this ideas with numerical analysis should be applicable to nonlinear acoustic experiments, opening new possibilities beyond the classical framework in deformation theory. For this reasons, it is necessary to be able to fix or anchor these models to well-established physical theories i.e. deformation theory. This provides, not only a formal basis to support such models, but also a series of rules to make solid and clear hypothesis 
in medicine. The possibility to build models from first principles allows us to have another critical perspective about the models per se, which may even be necessary to expand or modify the own foundations on which that models where deduced.  Any generalization of this case will allow not only the approach to new experimental situations but also a clearer way of characterizing materials in order to perform reverse engineering.


\section*{Acknowledgements}

This research was supported by the Ministry of Education DPI2017-83859- R, DPI2014-51870- R, and UNGR15-CE- 3664, and Junta de Andalucía PI-0107-2017, PIN-0030-2017.

\makeatletter
\makeatother

\begin{appendices}
\section{}\label{ApendA}


In the section \ref{AppA}, it has been showed that the elastic stress tensor is obtained from the elastic strain energy $U$ according to  the expression
\begin{equation}\label{first stress1}
\sigma_{ij}=\frac{\partial \,U}{\partial u_{i,j}}.
\end{equation}
If the elastic strain energy is given as a function $U(\pmb{\xi})$, there is an alternative way to compute the stress tensor  using the chain rule 
\begin{equation}\label{aca}
\frac{\partial \,U}{\partial u_{i,j}}=\mathlarger{\mathlarger{\sum}}_{k,l}\;\frac{\partial \,U}{\partial \xi_{kl}}\frac{\partial \xi_{kl}}{\partial u_{i,j}}.
\end{equation}
For nonlinear deformations, i.e. $\ell=1$ then $\xi_{ij}=\frac{1}{2}\big(u_{i,j}+u_{j,i}+u_{k,i}u_{k,j}\big)$, and 
$\sigma_{ij}$ from \eqref{first stress1} becomes 
\begin{equation}\label{strainstress1}
\sigma_{ij}=J_{ik}\frac{\partial \,U(\pmb{\xi})}{\partial\xi_{kj}}.
\end{equation} 
For linear deformations, i.e. $\ell=0$ then $\xi_{ij}=\frac{1}{2}\big(u_{i,j}+u_{j,i}\big)$,  and $\sigma_{ij}$ from \eqref{first stress1} becomes 
\begin{equation}\label{strainstress2}
\sigma_{ij}=\frac{\partial \,U(\pmb{\xi})}{\partial\xi_{kj}}.
\end{equation} 
The demonstration strategy of \eqref{strainstress1} and \eqref{strainstress2} consists in calculate $\partial \xi_{kl}/\partial u_{i,j}$. 

The calculation of $\partial \xi_{kl}/\partial u_{i,j}$ for $\ell=1$ is given that
\begin{align}\label{trabajo}
\frac{\partial \xi_{kl}}{\partial u_{i,j}}&=\frac{1}{2}\Bigg[\frac{\partial u_{k,l}}{\partial u_{i,j}}+\frac{\partial u_{l,k}}{\partial u_{i,j}}+\mathlarger{\mathlarger{\sum}}_n \frac{\partial u_{n,k}}{\partial u_{i,j}}u_{n,l}+u_{n,k}\frac{\partial u_{n,l}}{\partial u_{i,j}}\Bigg],\nonumber\\
\frac{\partial \xi_{kl}}{\partial u_{i,j}}&=\frac{1}{2}\Bigg[
\delta_{ki}\delta_{lj}+\delta_{li}\delta_{kj} +\mathlarger{\mathlarger{\sum}}_n \delta_{ni}\delta_{kj} u_{n,l}+u_{n,k}\delta_{ni}\delta_{lj}\Bigg],\nonumber\\
\frac{\partial \xi_{kl}}{\partial u_{i,j}}&=\frac{1}{2}\Big[\delta_{ki}\delta_{lj}+\delta_{li}\delta_{kj} +\delta_{kj} u_{i,l}+u_{i,k}\delta_{lj}\Big],\\
\frac{\partial \xi_{kl}}{\partial u_{i,j}}&=\frac{1}{2}\Big[(\delta_{ki}+u_{i,k})\delta_{lj}+ (\delta_{li}+ u_{i,l})\delta_{kj} \Big]\nonumber,\\
\frac{\partial \xi_{kl}}{\partial u_{i,j}}&=\frac{1}{2}\Big[J_{ik}\delta_{lj}+ J_{il}\delta_{kj} \Big].\nonumber
\end{align}
The conclusion of the last line is obtained using the definition of the displacement field $u_i=x'_i-x_i$, then $\delta_{ij}+u_{i,j}$ is equal to the jacobian matrix element $J_{ij}$. Replacing the last line in expression \eqref{trabajo} into \eqref{aca}
\begin{align}\label{aka}
\frac{\partial \,U}{\partial u_{i,j}}&= \frac{1}{2}\mathlarger{\mathlarger{\sum}}_{k,l}\;\frac{\partial \,U}{\partial \xi_{kl}}\bigg[J_{ik}\delta_{lj}+ J_{il}\delta_{kj} \bigg],\nonumber\\
\frac{\partial \,U}{\partial u_{i,j}}&= \frac{1}{2}
\mathlarger{\mathlarger{\sum}}_{k}\;\frac{\partial \,U}{\partial \xi_{kj}}J_{ik}+ \frac{1}{2}\mathlarger{\mathlarger{\sum}}_{l}\;\frac{\partial \,U}{\partial \xi_{jl}}J_{il}
,\\
\frac{\partial \,U}{\partial u_{i,j}}&= 
\mathlarger{\mathlarger{\sum}}_{k}\;J_{ik}\frac{\partial \,U}{\partial \xi_{kj}}
,\nonumber
\end{align}
where the symmetry of the strain tensor $\xi_{ij}=\xi_{ji}$ was used. The last line of \eqref{aka} complete the demonstration that \eqref{first stress1} and \eqref{strainstress1} are equivalent.

The calculation of $\partial \xi_{kl}/\partial u_{i,j}$ for $\ell=0$ is given that
\begin{equation}
\label{trabajo2}
\frac{\partial \xi_{kl}}{\partial u_{i,j}}=\frac{1}{2}\Big[\delta_{ki}\delta_{lj}+\delta_{li}\delta_{kj} \Big],
\end{equation}
also \eqref{aca} is significantly  reduced  to  
\begin{equation}\label{aca2}
\frac{\partial \,U}{\partial u_{i,j}}=\frac{\partial \,U}{\partial \xi_{ij}}
\end{equation}
both are equal to the elastic stress tensor $\sigma_{ij}$ in the linear case.

\section{}\label{ApendB}

Regarding the quadratic forms that appears in the dissipation function $\Psi$ and in the \textit{hookean} part of the potential energy $U$, its useful to prove the following identity for each $2-$rank and symmetric tensor $\pmb{\gamma}$
\begin{equation}\label{pirulo}
a \big(\gamma_{kk}\big)^2+b\Big(\gamma_{ij}-\frac{1}{3}\delta_{ij}\gamma_{kk}\Big)^2 =\left(a-\frac{1}{3}b\right) I_1^2(\pmb{\gamma})+b \,I_2(\pmb{\gamma})
\end{equation}
for any real values $(a,b)$.

In case that $\pmb{\gamma}=\pmb{\xi}$ or $\pmb{\gamma}=\partial_t\pmb{\xi}$, the left hand side of \eqref{pirulo} has a more easy physical interpretation. The first term contains the trace of $\pmb{\gamma}$ and the second term contains a traceless part of the tensor, then the parameter $a$ is associated with size changes and the parameter $b$ with shape changes. The right hand side of \eqref{pirulo} provides a more intuitive way to construct the potential energy and dissipative function, insomuch as under the isotropic hypothesis this two quantities depends only of the tensor invariants, and also it allows to calculate its derivatives using $\partial_{\pmb{\gamma}} I_n(\pmb{\gamma})=n\,\pmb{\gamma}^{n-1}$ \citep{lutkepohl},  in order to obtain the elastic and the dissipative stress tensors. 
Using the index summation convention $\gamma_{kk}=I_1(\pmb{\gamma})$ and $(\gamma_{ij})^2=\gamma_{ij}\gamma_{ji}=I_2(\pmb{\gamma})$, a demonstration of \eqref{pirulo} can be performed from its left member as follows 
\begin{align*}\label{pirulos}
a \big(\gamma_{kk}\big)^2+b\Big(\gamma_{ij}-\frac{1}{3}\delta_{ij}\gamma_{kk}\Big)^2 &=\cdots\nonumber\\
a  (\gamma_{kk})^2 + b\gamma_{ij}^2-\frac{2}{3}b\,\gamma_{ij}\delta_{ij}\gamma_{kk} +\frac{1}{9}b\,(\delta_{ij})^2(\gamma_{kk})^2 &=\cdots\nonumber\\
a  (\gamma_{kk})^2 + b\gamma_{ij}\gamma_{ji}-\frac{2}{3}b\,\gamma_{ii}\gamma_{kk} +\frac{1}{9}b\,\delta_{ij}\delta_{ji}(\gamma_{kk})^2 &=\cdots\nonumber\\
a I_1^2(\pmb{\gamma}) + bI_2(\pmb{\gamma})-\frac{2}{3}b\,I_1^2(\pmb{\gamma}) +\frac{1}{9}b\,I_2\left(\pmb{I}\right)I_1^2(\pmb{\gamma}) &=\cdots\nonumber\\
a I_1^2(\pmb{\gamma}) + bI_2(\pmb{\gamma})-\frac{2}{3}b\,I_1^2(\pmb{\gamma}) +\frac{1}{3}b\,I_1^2(\pmb{\gamma}) &=\cdots\nonumber\\
\left(a-\frac{1}{3}b\right) I_1^2(\pmb{\gamma})+b \,I_2(\pmb{\gamma})&=\cdots \square
\end{align*}
which is exactly equal to the right member of \eqref{pirulo}.

\section{}\label{ApendC}

Given a $2-$rank cartesian tensor $\pmb{\gamma}$ and a scalar function $F$ of its invariants $\{I_n(\pmb{\gamma}):=tr(\pmb{\gamma}^n) :\; n\in \mathbb{N}\}$, therefore the quantity  
\begin{equation}
\pmb{\varphi}=\partial_{\pmb{\gamma}} F 
\end{equation}
defines another $2-$rank cartesian tensor of whose components are $\varphi_{ij}=\partial_{\gamma_{ij}} F$, such that: 
\begin{equation}\label{aaa}
\partial_{\pmb{\gamma}} F=\sum_{n\geq 1} n\,\pmb{\gamma}^{n-1}\, \partial_{I_n} F,
\end{equation}
where the chain rule was used in \eqref{aca} and $\partial_{\pmb{\gamma}} I_n(\pmb{\gamma})=n\,\pmb{\gamma}^{n-1}$ \citep{lutkepohl}. Even in the case of $\gamma_{ij}=0$ except for $i=1=j$ (i.e. $\gamma_{ij}=\gamma_{11}\delta_{i1}\delta_{j1}$) then implies a nonzero components $\varphi_{22}$ and $\varphi_{33}$. The origin of this nonzero components comes from the dependence of $F$ with  $I_1(\pmb{\gamma})$:
\begin{equation}
\partial_{\pmb{\gamma}} F=\pmb{I}\partial_{I_1} F+2\pmb{\gamma}\partial_{I_2} F+3\pmb{\gamma}^2\partial_{I_3} F+\cdots
\end{equation}
where $\pmb{I}$ is the identity matrix. 

In summary, under the hypothesis $\gamma_{ij}=\gamma_{11}\delta_{i1}\delta_{j1}$ and $\pmb{\varphi}=\partial_{\pmb{\gamma}} F
$, if $F$ is not dependent on $I_1(\pmb{\gamma})$ then the components $\varphi_{22}$ and $\varphi_{33}$ are equal to zero.

The components of the elastic stress tensor $\pmb{\sigma}$, for the nonlinear case, and the dissipative stress tensor $\pmb{\sigma}'$ are explicitly obtained from \eqref{elastic stress} and \eqref{dissipative stress} as
\begin{align*}
\pmb{\sigma}=&\,\pmb{J}\big[\lambda I_1(\pmb{\xi})+\mathcal{B}I_2(\pmb{\xi})+\mathcal{C}I_1^2(\pmb{\xi})+\mathcal{E}I_3(\pmb{\xi})+2\mathcal{F}I_1(\pmb{\xi})I_2(\pmb{\xi})+  4\mathcal{H}I_1^3(\pmb{\xi})\big]\nonumber\\
&+2\pmb{J\xi}\big[\mu+\mathcal{B}I_1(\pmb{\xi})+\mathcal{F}I_1^2(\pmb{\xi})+2\mathcal{G}I_2(\pmb{\xi}) \big]+3\pmb{J\xi}^2\big[\tfrac{1}{3}\mathcal{A}+\mathcal{E}I_1(\pmb{\xi})\big],\\[15pt]
\pmb{\sigma}'=&2\eta\,\partial_t\,\pmb{\omega}+\left(\zeta+\chi-\frac{\,2\eta\,}{3}\right)I_1(\partial_t\,\pmb{\omega})\pmb{I}.
\end{align*}
The explicit form of each component of $\pmb{\sigma}$ and $\pmb{\sigma}'$ for $P-$waves in a one dimensional propagation  are given by
\begin{align*}
\sigma_{11}&=J_{11}\big[(\lambda+2\mu)\xi+(\mathcal{A+3B+C})\xi^2+2(\mathcal{E+2F+2G+2H})\xi^3\big],\nonumber\\
\sigma_{22}&=J_{22}\big[\lambda\xi+(\mathcal{B+C})\xi^2+2(\mathcal{E+F+2H})\xi^3\big],\nonumber\\
\sigma_{33}&=J_{33}\big[\lambda\xi+(\mathcal{B+C})\xi^2+2(\mathcal{E+F+2H})\xi^3\big],\nonumber\\
\sigma_{11}'&=\left(\frac{\,4\eta\,}{3}+\zeta+\chi\right)\partial_t(\partial_x u),\nonumber\\
\sigma_{22}'&=\left(\zeta+\chi-\frac{\,2\eta\,}{3}\right)\partial_t(\partial_x u),\nonumber\\
\sigma_{33}'&=\left(\zeta+\chi-\frac{\,2\eta\,}{3}\right)\partial_t(\partial_x u).
\end{align*}
where by definition $J_{ij}=\delta_{ij}+u_{i,j}$, and for this case: $J_{11}=1+\partial_x u$ and $J_{22}=1=J_{33}$.

In order to obtain  $\sigma_{11}=\sigma_{22}=\sigma_{33}$  and 
$\sigma'_{11}=\sigma'_{22}=\sigma'_{33}$ the shear modulus $\mu$ and shear viscosity $\eta$ must be neglected, necessarily, because only \textit{compression} forces can act on the body, this condition simulate the fluid behaviour. From a linear deformation $\ell=0$, if $\mu=0$ and $\eta=0$ the hydrostatic pressure in solids or fluids. share the same definition. From a nonlinear deformation $\ell=1$,  if $\mu=0$ and $\eta=0$ the solid behaviour is quite similar to fluid behaviour, because the differences between the diagonal elements of $\pmb{\sigma}$ are small: $\sigma_{ii}-\sigma_{jj} = \mathcal{O}(\pmb{\xi}^2)$, for $i\neq j$. The absence of higher order terms in $\pmb{\sigma}'$ allows that $\eta=0$ implies $\sigma'_{ii}=\sigma'_{jj}$, even in the case of nonlinear deformations. 

Taking the general expression of the $i-$component of the force per unit of volume is given by $f_i=\partial_{x_j}\sigma_{ij}$. The force per unit of volume is a vector field along the $x-$direction. For this reason the relevant component of the total stress tensor that has been considered in this work was $\sigma_{11} + \sigma_{11}'$.

\end{appendices}

\bibliography{bib}

\begin{thebibliography}{10}
\providecommand{\url}[1]{{#1}}
\providecommand{\urlprefix}{URL }
\providecommand{\doi}[1]{\url{https://doi.org/#1}}
\bibcommenthead

\bibitem{Westervelt63}
P.~Westervelt, Parametric acoustic array.
\newblock The Journal of the Acoustical Society of America \textbf{35},
  535--537 (1963)

\bibitem{landau86}
L.D. Landau, E.M. Lifshitz, \emph{Theory of Elasticity} (Pergamon, Oxford, USA,
  1986)

\bibitem{Marsden}
J.E. Marsden, T.~Hughes, \emph{Mathematical Foundations of Elasticity} (Dover,
  USA, 1983)

\bibitem{hamilton1998nonlinear}
M.F. Hamilton, D.T. Blackstock, et~al., \emph{Nonlinear acoustics}, vol.~1
  (Academic press San Diego, 1998)

\bibitem{Jimenez}
N.~Jim\'enez, Nonlinear acoustic waves in complex media.
\newblock Dissertation, Universitat Polit\'ecnica de Valencia (2015)

\bibitem{SHEVCHENKO2015200}
Absorbing boundary conditions for nonlinear acoustics: The westervelt equation.
\newblock J. Comp. Phys. \textbf{302}, 200 -- 221 (2015)

\bibitem{Pernot02}
M.~Pernot, K.~Waters, J.~Berco, M.~Tanter, M.~Fink, Reduction of the
  thermoacoustic lens effect during ultrasound based temperature estimation.
\newblock Proceedings of 2002, IEEE Ultrasonics Symposium pp. 1447--1450 (2002)

\bibitem{Simon98}
C.H. Scholz, S.H. Hickman, Two-dimensional temperature estimation using
  diagnostic ultrasound.
\newblock IEEE Trans. Ultrason. Ferr. \textbf{45}(4), 1088--1099 (1998)

\bibitem{Floch97}
C.~Le~Floch, M.~Fink, Ultrasonic mapping of temperature in hyperthermia: the
  thermal lens effect.
\newblock Proceedings 1997, IEEE Ultrasonics Symposium pp. 1301--1304 (1997)

\bibitem{Connor02}
C.~Connor, K.~Hynynen, Bio-acoustic thermal lensing and nonlinear propagation
  in focused ultrasound surgery using large focal spots: a parametric study.
\newblock Journal of the Acoustical Society of America \textbf{47}(11),
  1911--1928 (2002)

\bibitem{Hallaj01}
I.M. Hallaj, R.~Cleveland, K.~Hynynen, Simulations of the thermo-acoustic lens
  effect during focused ultrasound surgery.
\newblock J. Acoust. Soc. Am. \textbf{109}(5), 2245--2253 (2001)

\bibitem{Floch99}
C.~Le~Floch, M.~Tanter, M.~Fink, Self-defocusing in ultrasonic hyperthermia:
  Experiment and simulation.
\newblock Proceedings 1997, IEEE Ultrasonics Symposium \textbf{74}(20),
  3062--3064 (1999)

\bibitem{clason2009boundary}
C.~Clason, B.~Kaltenbacher, S.~Veljovi{\'c}, Boundary optimal control of the
  westervelt and the kuznetsov equations.
\newblock Journal of Mathematical Analysis and Applications \textbf{356}(2),
  738--751 (2009)

\bibitem{taraldsen2001generalized}
G.~Taraldsen, A generalized westervelt equation for nonlinear medical
  ultrasound.
\newblock The Journal of the Acoustical Society of America \textbf{109}(4),
  1329--1333 (2001)

\bibitem{Averkiou99}
M.~Averkiou, R.~Cleveland, Modeling of an electrohydraulic lithotripter with
  the kzk equation.
\newblock Journal of the Acoustical Society of America \textbf{106}(1),
  102--112 (1999)

\bibitem{Kaltenbacher07}
M.~Kaltenbacher, \emph{Numerical simulation of mechatronic sensors and
  actuators} (1999)

\bibitem{Dreyer00}
T.~Dreyer, W.~Kraus, E.~Bauer, R.E. Riedlinger, Investigations of compact self
  focusing transducers using stacked piezoelectric elements for strong sound
  pulses in therapy.
\newblock In Proceedings of the IEEE Ultrasonics Symposium pp. 1239--1242
  (2000)

\bibitem{karamalis2010fast}
A.~Karamalis, W.~Wein, N.~Navab, in \emph{Medical Image Computing and
  Computer-Assisted Intervention--MICCAI 2010} (Springer, 2010), pp. 243--250

\bibitem{varray2011simulation}
F.~Varray, C.~Cachard, A.~Ramalli, P.~Tortoli, O.~Basset, Simulation of
  ultrasound nonlinear propagation on gpu using a generalized angular spectrum
  method.
\newblock EURASIP journal on Image and Video Processing \textbf{2011}(1), 1--6
  (2011)

\bibitem{solovchuk2012effects}
M.A. Solovchuk, T.W. Sheu, M.~Thiriet, in \emph{12th International Symposium on
  Therapeutic Ultrasound}, vol. 1503 (AIP Publishing, 2012), pp. 83--88

\bibitem{xiaorui2011simulation}
C.~Xiaorui, Z.~Xiaojing, W.~Shaolin, J.~Xiqi, in \emph{Biomedical Engineering
  and Informatics (BMEI), 2011 4th International Conference on}, vol.~2 (IEEE,
  2011), pp. 986--989

\bibitem{thiriet2014hifu}
M.~Thiriet, M.~Solovchuk, T.W.H. Sheu, in \emph{Computational Modeling of
  Objects Presented in Images. Fundamentals, Methods, and Applications}
  (Springer, 2014), pp. 1--11

\bibitem{Bochud}
N.~Bochud, Signal processing-based identification of pathology using
  ultrasonics.
\newblock Dissertation, Universidad de Granada (2014)

\bibitem{ma12040607}
W.~Lyu, X.~Wu, W.~Xu, Nonlinear acoustic modeling and measurements during the
  fatigue process in metals.
\newblock Materials \textbf{12}(4) (2019)

\bibitem{guyer99b}
R.~Guyer, P.~Johnson, Nonlinear mesoscopic elasticity: evidence for a new class
  of materials.
\newblock Physics Today \textbf{52}(4), 30--36 (1999)

\bibitem{GIAMMARINARO2018236}
B.~Giammarinaro, A.~Zorgani, S.~Catheline, Shear-wave sources for soft tissues
  in ultrasound elastography.
\newblock IRBM \textbf{39}(4), 236 -- 242 (2018)

\bibitem{donskoy2}
D.~Donskoy, A.~Sutin, A.~Ekimov, Nonlinear acoustic interaction on contact
  interfaces and its use for nondestructive testing.
\newblock Ndt \& E International \textbf{34}(4), 231--238 (2001)

\bibitem{Sutin1995}
A.M. Sutin, V.E. Nazarov, Nonlinear acoustic methods of crack diagnostics.
\newblock Radiophysics and Quantum Electronics \textbf{38}(3), 109--120 (1995)

\bibitem{Zarembo1989}
L.K. Zarembo, V.A. Krasil'nikov, I.E. Shkol'nik, Nonlinear acoustics in a
  problem of diagnosing the strength of solids.
\newblock Strength of Materials \textbf{21}(11), 1544--1551 (1989)

\bibitem{landau80}
L.D. Landau, E.M. Lifshitz, \emph{Statistical Physics} (Pergamon, Oxford, USA,
  1980)

\bibitem{norris1998finite}
A.~Norris, \emph{Nonlinear acoustics} (Academic Press, San Diego, 1998), chap.
  Finite amplitude waves in solids, pp. 263--277

\bibitem{murnaghanfinite}
F.~Murnaghan, \emph{Finite deformation of an elastic solid} (Chapman \& Hall,
  London, 1951)

\bibitem{lutkepohl}
H.~Lutkepohl, \emph{Handbook of Matrices} (John Wiley \& Sons, USA, 1996)

\bibitem{Frobenius}
G.~Frobenius, Ueber lineare substutionen und bilineare formen.
\newblock J. Reine Angew. Math. \textbf{84}, 1--63 (1878)

\bibitem{hamilton2004nonlinear}
Y.A. Hamilton Mark F.;~Ilinskii, E.A. Zabolotskaya, \emph{Separation of
  compressibility and shear deformation in the elastic energy density (L)},
  vol.~1 (Journal of the Acoustical Society of America 116, 41, 2004)

\bibitem{Mohanty51}
S.R. Mohanty, A relationship between heat conductivity and viscosity of
  liquids.
\newblock Nature \textbf{168}(4262), 42--42 (1951)

\bibitem{O'Neal62}
C.~O'Neal, R.S. Brokaw, Relation between thermal conductivity and viscosity for
  some nonpolar gases.
\newblock The Physics of Fluids \textbf{5}(5), 567--574 (1962)

\bibitem{landau87}
L.D. Landau, E.M. Lifshitz, \emph{Fluid Dynamics} (Pergamon, Oxford, USA, 1987)

\bibitem{polyanin2003handbook}
A.D. Polyanin, V.F. Zaitsev, \emph{Handbook of nonlinear partial differential
  equations} (CRC press, 2003)

\bibitem{Bachmann}
P.~Bachmann, \emph{Analytische Zahlentheorie}, vol.~2 (Leipzig, 1894)

\bibitem{LandauE}
E.~Landau, \emph{Handbuch der Lehre von der Verteilung der Primzahlen}
  (Leipzig, USA, 1909)

\bibitem{Morse}
P.M. Morse, H.~Feshbach, \emph{Methods of Theoretical Physics}, vol.~I
  (McGraw-Hill, pp. 411-413, 1953)

\bibitem{Abramowitz}
M.~Abramowitz, I.A.E. Stegun, \emph{Handbook of Mathematical Functions with
  Formulas, Graphs, and Mathematical Tables} (Dover, p. 16, 1972)

\bibitem{Kuznetsov71}
V.~Kuznetsov, Equations of nonlinear acoustics.
\newblock Sov. Phys. Acous. \textbf{16}, 467--470 (1971)

\bibitem{zabolotskaya69}
E.~Zabolotskaya, R.~Khokhlov, Quasi-plane waves in the nonlinear acoustics of
  confined beams.
\newblock Sov. Phys. Acoust \textbf{15}(1), 35--40 (1969)

\bibitem{Meyer2011}
S.~Meyer, M.~Wilke, Optimal regularity and long-time behavior of solutions for
  the westervelt equation.
\newblock Applied Mathematics {\&} Optimization \textbf{64}(2), 257--271 (2011)

\bibitem{Kaltenbacher2011}
B.~Kaltenbacher, I.~Lasiecka, S.~Veljovi{\'{c}}, \emph{Well-posedness and
  Exponential Decay for the Westervelt Equation with Inhomogeneous Dirichlet
  Boundary Data} (Springer Basel, Basel, 2011), pp. 357--387

\end{thebibliography}

\end{document}